\documentclass[12pt,nofootinbib,aps,amsmath,amssymb,preprint,showpacs]{revtex4}
\usepackage[T1]{fontenc} \usepackage[latin1]{inputenc}
\usepackage{amsmath,color,ulem} \usepackage{graphicx}
\usepackage{epsfig} \usepackage{amssymb}
\usepackage{amsmath,amssymb,graphicx,amssymb,color,cancel}
\usepackage{ulem,cancel}

\makeatletter 

\begin{document}

\title{Efficient simulation of semiflexible polymers}

\author{Debabrata Panja} \affiliation{Institute for Theoretical
  Physics, Universiteit Utrecht, Leuvenlaan 4, 3584 CE Utrecht, The
  Netherlands} \affiliation{Institute of Physics, Universiteit van
  Amsterdam, Science Park 904, Postbus 94485, 1090 GL Amsterdam, The
  Netherlands} \author{Gerard T. Barkema} \affiliation{Institute for
  Theoretical Physics, Universiteit Utrecht, Leuvenlaan 4, 3584 CE
  Utrecht, The Netherlands} \affiliation{Instituut-Lorentz,
  Universiteit Leiden, Niels Bohrweg 2, 2333 CA Leiden, The
  Netherlands} \author{J.M.J. van Leeuwen}
\affiliation{Instituut-Lorentz, Universiteit Leiden, Niels Bohrweg 2,
  2333 CA Leiden, The Netherlands}

\begin{abstract} Using a recently developed bead-spring model for
  semiflexible polymers that takes into account their natural
  extensibility, we report an efficient algorithm to simulate the
  dynamics for polymers like double-stranded DNA (dsDNA) in the
  absence of hydrodynamic interactions. The dsDNA is modelled with one
  bead-spring element per basepair, and the polymer dynamics is
  described by the Langevin equation. The key to efficiency is that we
  describe the equations of motion for the polymer in terms of the
  amplitudes of the polymer's fluctuation modes, as opposed to the use
  of the physical positions of the beads. We show that, within an
  accuracy tolerance level of $5\%$ of several key observables, the
  model allows for single Langevin time steps of $\approx1.6$, 8, 16
  and 16 ps for a dsDNA model-chain consisting of 64, 128, 256 and 512
  basepairs (i.e., chains of 0.55, 1.11, 2.24 and 4.48 persistence
  lengths) respectively. Correspondingly, in one hour, a standard
  desktop computer can simulate 0.23, 0.56, 0.56 and 0.26 ms of these
  dsDNA chains respectively. We compare our results to those obtained
  from other methods, in particular, the (inextensible discretised)
  WLC model. Importantly, we demonstrate that at the same level of
  discretisation, i.e., when each discretisation element is one
  basepair long, our algorithm gains about 5-6 orders of magnitude in
  the size of time steps over the inextensible WLC model. Further, we
  show that our model can be mapped one-on-one to a discretised
  version of the extensible WLC model; implying that the speed-up we
  achieve in our model must hold equally well for the latter. We also
  demonstrate the use of the method by simulating efficiently the
  tumbling behaviour of a dsDNA segment in a shear flow.
\end{abstract}

\pacs{36.20.-r,64.70.km,82.35.Lr}

\maketitle

\section{Introduction\label{sec1}}

Over the last decades, there has been a surge in research activities
in the physical properties of biopolymers, such as double-stranded DNA
(dsDNA), filamental actin (F-actin) and microtubules. Semiflexibility
is a common feature they share: they preserve mechanical rigidity over
a range, characterised by the persistence length $l_p$, along their
contour. (E.g., for a dsDNA, F-actin and microtubules, $l_p\approx40$
nm \cite{dnapersist,wang}, $\sim16 \mu$m \cite{Factinpersist} and
$\sim5$ mm \cite{micropersist} respectively.) 

Recently, two of us introduced a bead-spring model for semiflexible
polymers \cite{leeuwen}. The model has four parameters. Three of them
determine the mechanical properties: the average inter-bead
distance $a$, the longitudinal stiffness $\lambda$, and the bending
stiffness $\kappa$. The fourth parameter is related to the viscosity
of water $\xi$ and sets the time scale. In earlier work
\cite{leeuwen,leeuwen1} we determined a set of values for the first
three parameters that are able to reproduce the mechanical properties
of double-stranded DNA in experiment, and we studied the canonical averages 
of a number of equilibrium properties. In most of the present paper, we
restrict ourselves to the same set of parameter values.

In this paper, we primarily discuss how the equations of motion of our
model can be efficiently integrated in time.  There is no hydrodynamic
interactions among the beads. This is in fact not a problem when it
comes to comparing to experimental results for dsDNA segments up to a
few persistence lengths, since at these lengths a semiflexible polymer
does not form a coil, and therefore should be free-draining. Indeed,
this is the feature that allows us to meaningfully compare the
diffusion coefficients of short model dsDNA segments to those from
experiments, from which we determine the fourth (and the last)
parameter of the model, $\xi$, which describes the Langevin friction
on the beads.

The default simulation approach to integrate the corresponding
Langevin equations of motion in time for our model would be a simple
integration scheme such as the Euler method, using the bead positions
as dynamical variables. In this paper we develop a time-forward
integration scheme by using the properties of (a very good
approximation of) the polymer's fluctuation modes in this model
\cite{leeuwen1}, and allowing a set of representative equilibrium and
dynamical observables to differ by at most 5\%, we achieve 2-3 orders
of magnitude speed-ups in comparison to the default method. With
average inter-bead distance $a\approx0.33$ nm as a model parameter,
the length of a dsDNA basepair, the maximum size $\Delta
t_{\text{max}}$ of the time step is summarised in Table \ref{table0}.
\begin{table*}
\begin{center}
  \begin{tabular}{c||c} \hline\hline$\quad$ chain length
    (bp)$\quad$ &$\quad\Delta t_{\text{max}}$ (ps) $\quad$ 
    \tabularnewline\hline\hline 64&1.59\tabularnewline\hline
    128&7.96\tabularnewline\hline
    256&15.9\tabularnewline\hline
    512&15.9\tabularnewline\hline\hline
\end{tabular}
\end{center}
\caption{With average inter-bead distance $a\approx0.33$ nm --- the
  length of a dsDNA basepair --- as a model
  parameter, the maximum size of the integration time step $\Delta
  t_{\text{max}}$ are shown for various chain lengths for dsDNA. The
  persistence length we use is $l_p\approx37.6$ nm \cite{wang},
  corresponding to $\approx 114$ basepairs. \label{table0}}
\end{table*}

We also relate our work to existing theoretical work on semiflexible
polymers. We show that our model can be mapped one-on-one to the
discretised version of the extensible wormlike chain (WLC) model
\cite{ober}; implying that the speed-up we achieve in our model must
hold equally well for the latter. We further demonstrate that if in
our model the longitudinal stiffness $\lambda$ is made very large
while keeping fixed its resistance to bending $\kappa$, it effectively
reduces to a discretised version of the inextensible WLC model
\cite{kratky}. (The inextensible WLC model, its subsequent
modifications \cite{mods1,mods2,mods3,mods4}, and recent analyses
\cite{recent1,recent2, recent3, recent4, recent5, recent6} have been
very successful in describing static/mechanical properties of dsDNA,
such as its force-extension curve and the radial distribution function
of its end-to-end distance.) Since the contour length of the polymer
is constrained in the original inextensible WLC model, Lagrangian
multipliers of a varying degree of sophistication have been introduced
in its computer implementation in order to enforce a contour length
that is either strictly fixed
\cite{liverpool1,liverpool2,liverpool3,liverpool4,kroy,morse,montesi},
or fixed on average \cite{winkler1,winkler2}. In particular, we show
that in the limit of large $\lambda$ at fixed $\kappa$ the dynamical
equations of the beads in our model approach those similar to the ones
that Morse \cite{morse,montesi} developed in order to simulate the
inextensible WLC. Using this relation between our model and the
inextensible WLC, we demonstrate that the maximal allowable time step
$\Delta t_{\text{max}}$ for the inextensible WLC model of a dsDNA
chain of 63 basepairs is $\approx 0.02$ fs, in good agreement with
Ref. \cite{frey} (that has recently implemented Morse's algorithm for
the inextensible WLC). In other words, as shown in Table \ref{table0},
to simulate dsDNA {\it our model achieves a maximal allowable time
  step that is 5-6 orders of magnitude larger than that of the
  inextensible WLC.}

This paper is organised as follows. In Sec. \ref{sec2} we briefly
introduce the model, and identify the parameter values of the model
Hamiltonian for dsDNA. Here we also show that our model, in the
parameter space, can be mapped one-on-one to the discretised version
of the extensible WLC model. In Sec. \ref{sec3} we describe the
equations for polymer dynamics. Section \ref{sec4} is devoted to the
time-integrated algorithm for the equation of motion for the polymer
in mode representation. In Sec. \ref{sec5} we test the
time-integration algorithm for dsDNA. In Sec. \ref{newsec} we discuss
coarse-graining in our model, which leads us to the result that in the
limit of large $\lambda$ at fixed $\kappa$ the dynamical equations of
the beads in our model approach those similar to the ones that Morse
\cite{morse,montesi} developed in order to simulate the inextensible
WLC.  In Sec. \ref{sec6} we elaborate on our numerical results
presented in Table \ref{table0}, and we conclude the paper with a
discussion in Sec. \ref{sec7}, including a wider comparison to the
time steps achieved in the existing literature. Finally, in the
Supplementary Information we present a movie of a tumbling dsDNA 
segment in a shear flow, generated by the use of this algorithm, to 
illustrate the usefulness of the simulation approach.

\section{The model\label{sec2}}

The model we use for semiflexible polymers is described in detail in
Ref. \cite{leeuwen}. The Hamiltonian for the model is of the form
\begin{eqnarray}
{\cal H} =\frac\lambda2 \sum^N_{n=1} (|{\bf u}_n|-d)^2 - \kappa \sum^{N-1}_{n=1} {\bf u}_n \cdot {\bf u}_{n+1}.
\label{a1}
\end{eqnarray} 
Here ${\bf u}_n = {\bf r}_n - {\bf r}_{n-1}$ is the bond vector
between bead $n-1$ and $n$, with ${\bf r}_n$ being the position of the
$n$-th bead ($n=0,1, \cdots , N$).  The first term in this Hamiltonian
relates to the longitudinal stiffness of the chain, while the second
term relates to its resistance to bending. The parameters in the
Hamiltonian are the following. The quantity $d$ sets the length scale
of a bond --- if only the first term would be present, bonds would
assume the length $d$. The presence of the second term in the
Hamiltonian causes an elongation of the bonds such that the average
bond length is $a=bd$, with a factor $b$ that depends on the type of
the polymer. Further, $\lambda$ and $\kappa$ are two parameters,
relating to the longitudinal (stretching) and transverse (bending)
stiffness of the chain. In order to have ${\cal H}$ represent
semiflexible polymers, both parameters $\lambda$ and $\kappa$
typically will have to be large. Instead of working in terms of
$\lambda$ and $\kappa$, we choose the ratios $\nu=\kappa /\lambda$ and
$T^*=k_BT/(\lambda d^2)$ as characteristic parameters to describe the
model \cite{leeuwen}, which reduces the Hamiltonian to
\begin{eqnarray} \frac{\cal H}{k_BT}=\frac{1}{2T^*} \left[\sum^N_{n=1}
(u_n-1)^2 - 2\nu \sum^{N-1}_{n=1} {\bf u}_n \cdot {\bf
u}_{n+1}\right],
\label{a2}
\end{eqnarray} 
with $u_n \equiv \left| {\bf u}_n\right|$ is the length of the scaled bond vector.
Note that stability of the Hamiltonian requires
$0<\nu< 1/2$, and that in these variables \cite{leeuwen,leeuwen1}
\begin{eqnarray} 
b=\frac{1}{1-2\nu}\quad\mbox{and}\quad l_p=\frac{ab^2\nu}{T^*}=\frac{\kappa a^3}{k_BT}.
\label{a3}
\end{eqnarray}

\subsection{The model parameters for dsDNA\label{sec2a}}

For dsDNA the physical distance between
the beads equals $a=0.33$ nm, i.e. the length of a basepair.  The two
parameters $T^*$ and $\nu$ can be chosen by matching the
force-extension curve for the polymer, leading to $\nu=0.35$ and
$T^*=0.034$ \cite{leeuwen}. Following Eq.(\ref{a3}), the factor $b$
then turns out to be $\approx 3.3$, so that the length parameter
$d\approx0.1$ nm.  The number of beads
$(N+1)$ simply equals the number of basepairs present in the dsDNA
chain.

The equilibrium and the dynamical properties of the model, specially
in relation to the well-known properties of semiflexible polymers have
been studied in detail in Refs. \cite{leeuwen,leeuwen1}. Nevertheless,
in order to demonstrate the usability of this model for reaching long
length and time-scales on a computer we need to revisit the dynamical
equations resulting from the Hamiltonian (\ref{a2}).

\subsection{Relating our model to the extensible WLC\label{newseca}}

We start with the expression for the extensible WLC as given by
Obermayer and Frey \cite{ober}
\begin{equation} 
  {\cal H} = \frac{k_B T}{2} \int^L_0 ds
  \left(l_p\, |{\bf r}''|^2 + k_x\, [|{\bf r}'| -1]^2 \right),
  \label{w1} 
\end{equation} 
where ${\bf r} (s)$ is the contour of the chain, ${\bf r}'$ the first
derivative and ${\bf r}''$ the second derivative with respect to the
contour length parameter $s$. In order to simulate the dynamical
behaviour of the continuous chain, the chain is represented by a set of
$N$ discrete points
\begin{equation} 
  s_n = n \Delta s, \quad \quad \quad L =N
  \Delta s, \quad \quad \quad {\bf r}_n = {\bf r} (s_n).
  \label{w2} 
\end{equation} 
The derivatives are replaced by
\begin{equation} 
  {\bf r}' \Rightarrow \frac{{\bf r}_n -
    {\bf r}_{n-1} }{\Delta s} = \frac{{\bf u}_n}{\Delta s},
  \label{w3} 
\end{equation} 
and
\begin{equation} 
  {\bf r}'' \Rightarrow \frac{{\bf u}_{n+1}
    - {\bf u}_n }{\Delta s} = \frac{{\bf r}_{n+1}- 2 {\bf r}_n +{\bf
      r}_{n-1}} {(\Delta s)^2}
  \label{w4} 
\end{equation} 
The points on the chain will correspond to the beads of our
Hamiltonian.  We take $\Delta s = a$ as the distance between the
points for a transparent comparison between the models.

Inserting these derivatives into the Hamiltonian (\ref{w1}) yields
\begin{equation} 
  {\cal H} = \frac{k_B T}{2 a} \sum^N_{n=1}
  \left( l_p |{\bf u}_{n+1} - {\bf u}_n|^2 + k_x [u_n -a]^2 \right).
  \label{w5} 
\end{equation} 
Writing out the squares and collecting the terms of the same nature we
get
\begin{equation} 
  {\cal H} = \frac{k_B T}{2 a} \sum_n \left(
    [2 l_p +k_x] u^2_n - 2 l_p {\bf u}_n \cdot {\bf u}_{n+1} - 2 k_x a \,
    u_n \right).
  \label{w6} 
\end{equation} 
We have left out the irrelevant constant and ignored the minor difference between 
the coefficient of first and last bond in the term with $u^2_n$ and those of 
the other bonds.

In order to compare the expression (\ref{w6}) with our Hamiltonian (\ref{a1}) we
must realise that the Hamiltonian (\ref{w1}) uses a scaling that is
not the same as ours. So there is an overall constant $f$ difference
between the two Hamiltonians. Keeping this in mind we get the
relations
\begin{equation} 
  f \, \frac{k_b T}{a} l_p = \kappa, \quad
  \quad f\, \frac{k_b T}{2 a} (2 l_p +k_x) = \lambda, \quad \quad f \,
  k_b T\, k_x = \lambda d.
  \label{w7} 
\end{equation} 
The overall factor $f$ is determined from the second relation (\ref{a3}) between the 
persistence length $l_p$ and our constant $\kappa$. The first relation (\ref{w7}) yields
\begin{equation} 
  f = 1/a^2.
  \label{w9} 
\end{equation} 
Using this in the last relation of Eq. (\ref{w7}) we get the connection between $k_x$ and $\lambda$.
\begin{equation} 
  k_x = \frac{\lambda d a^2}{k_B T}.
  \label{w10} 
\end{equation} 
Inserting Eqs. (\ref{a3}) and (\ref{w10}) into the middle relation
of Eq. (\ref{w7}) leads to the relation
\begin{equation} 
  2 \kappa + \frac{\lambda d}{a} = \lambda,
  \quad \quad {\rm or} \quad \quad \frac{d}{a} =\frac{1}{b} = 1 - 2 \nu,
\label{w11} 
\end{equation}
which is consistent with the first relation  (\ref{a3}).

Thus the discretised extensible WLC is identical to our model with the
above given connection of the parameters, except for a small
difference for the strength of the interaction parameters of the first
and last bond. This implies that any conclusion we draw on our model
is equally valid for discretised versions of the extensible WLC.

\section{Polymer dynamics \label{sec3}}

We describe the polymer dynamics in terms of the Langevin equation. It
is natural to choose the positions of the beads as the dynamical
variables, obeying the equations
\begin{eqnarray} \frac{d {\bf r}_n (t)}{d t } = -\frac1\xi\frac{\partial
{\cal H}}{\partial {\bf r}_n} + {\bf g}_n (t).
\label{b1}
\end{eqnarray} Here $\xi$ is the friction coefficient and ${\bf g}_n$
is the Gaussian distributed random thermal force on bead $n$ due to
the solvent molecules, with the fluctuation spectrum
\begin{eqnarray} \langle g^\alpha_m (t)\, g^\beta_n (t') \rangle =
\frac{2 k_B T}{\xi} \, \delta^{\alpha,\beta} \, \delta_{m,n} \,
\delta(t-t').
\label{b2}
\end{eqnarray} 
For numerical evaluation of these equations it is useful to reduce
time and distances to dimensionless variables. We therefore scale
distances and time by $d$ and time by $\xi/\lambda$, i.e., we write
the bead positions as ${\bf r}_n={\bf r}'_nd$ and $t
=\xi\tau/\lambda$, which gives the Langevin equation the form
\begin{eqnarray} \frac{d {\bf r}'_n (t)}{d \tau } = - \frac{\partial
{\cal H}'}{\partial {\bf r}'_n} + {\bf g}'_n (\tau).
\label{b3}
\end{eqnarray} Correspondingly, the dimensionless random force
\begin{eqnarray} {\bf g}'_n = \frac{{\bf g}_n d}{\lambda}
\label{b4}
\end{eqnarray} 
has the correlation function
\begin{eqnarray} \langle g'^\alpha_m (\tau)\, g'^\beta_n (\tau')
\rangle = {2 T^*} \, \delta^{\alpha,\beta} \, \delta_{m,n} \,
\delta(\tau-\tau').
\label{b5}
\end{eqnarray} In order to restore notational simplicity henceforth we
omit the primes on the variables.

\subsection{The dynamical equations in terms of polymer's fluctuation
modes\label{sec3a}}

It is of course possible to simulate polymer dynamics using the
default Euler method, Eqs. (\ref{b3}-\ref{b5}), with the bead
positions as variables. This however only allows Langevin time step
$\Delta\tau=0.1$, and at $\Delta\tau\approx0.3$ (corresponding to 0.16
and 0.48 ps respectively --- $\Delta\tau=1$ corresponds to 1.5 ps, see
Sec. \ref{sec7b}) the integration scheme even becomes unstable. An
equivalent manner to simulate polymer dynamics is to use its
fluctuation modes as variables. The main advantage of the latter is
that the modes with longer length-scales have slower decay times, and
as result one can make a separation in time scales, which in turn
allows for the possibility of larger time steps, i.e., faster
simulations that eventually achieves 2-3 orders of magnitude larger
integration time steps. In this section we describe the method.

As for describing polymer dynamics in terms of the polymer's
fluctuation modes (described by the mode variables ${\bf R}_p$), note
that any transformation of the type
\begin{eqnarray} \left\{ \begin{array}{rcl} {\bf R}_p & = &
\displaystyle \sum_n {\bf r}_n \, \phi_{n,p}, \\*[4mm] {\bf r}_n & = &
\displaystyle \sum_p \phi_{n,p}\, {\bf R}_p,
\end{array} \right.
\label{c1}
\end{eqnarray} where $\phi_{n,p}$ is an orthogonal matrix, satisfying
\begin{eqnarray} \sum_p \phi_{m,p}\, \phi_{n,p} =\delta_{mn},
\label{c2}
\end{eqnarray} leaves the dynamical equation (\ref{b3}) form
invariant; i.e.,
\begin{eqnarray} \frac{d {\bf R}_p (t)}{d \tau } = - \frac{\partial
{\cal H}}{\partial {\bf R}_p} + {\bf G}_p.
\label{c3}
\end{eqnarray} Here ${\bf G}_p$ is the transform of ${\bf g}_n$:
\begin{eqnarray} {\bf G}_p = \sum_n {\bf g}_n \, \phi_{n,p},
\label{c4}
\end{eqnarray} whereas the derivative with respect to ${\bf R}_p$ can
be calculated with the chain rule
\begin{eqnarray} \frac{\partial {\cal H}}{\partial {\bf R}_p} = \sum_n
\frac{\partial {\cal H}} {\partial {\bf r}_n} \phi_{n,p}.
\label{c5}
\end{eqnarray}

Returning to our Hamiltonian (\ref{a2}), we see that it can be
rewritten in the form \cite{leeuwen}
\begin{eqnarray} \frac{{\cal H}}{k_BT}-N/2 = \frac{1}{2} \sum_{m,n}
{\bf r}_m \cdot H_{m,n} {\bf r}_n - L_c={\cal H}^*- L_c,
\label{c6}
\end{eqnarray} with $L_c$ the contour length
\begin{eqnarray} L_c = \sum_n u_n.
\label{c7}
\end{eqnarray} In this form of the Hamiltonian, the ${\cal H}^*$ term
is not only quadratic in the bead positions, but also $H_{mn}$ becomes
diagonal under the transformation ($p=0,1,\ldots,N-1$) \cite{leeuwen}
\begin{eqnarray} \phi_{n,p} = \left(\frac{2}{N+1}\right)^{1/2} \, \cos
\left( \frac{p(n+1/2)\pi}{N+1} \right),
\label{c8}
\end{eqnarray} which are of the same form as the Rouse modes for a
flexible polymer \cite{rouse}, with eigenvalues
\begin{eqnarray} \zeta^l_p = 2\left[1- \cos \left(\frac{p\pi}{N+1}
\right)\right] \left[1- 2 \nu \cos \left(\frac{p\pi}{N+1}
\right)\right].
\label{c9}
\end{eqnarray} In other words, ${\cal H}^*$ is simply expressed as
\begin{eqnarray} {\cal H}^*=\frac12\sum_p\zeta^l_pR^2_p.
\label{c10}
\end{eqnarray}

Unfortunately though, the term $L_c$ in the Hamiltonian (\ref{c6}) is
not diagonal in the Rouse mode representation, meaning that $L_c$
contains coupling among different Rouse modes. Consequently, the
equation of motion for the polymer takes the form
\begin{eqnarray} \frac{d {\bf R}_p (t)}{d \tau } = -\zeta_p^l {\bf
R}_p+{\bf H}_p + {\bf G}_p,
\label{c11}
\end{eqnarray} where ${\bf H}_p=-\partial{L_c}/\partial{\bf
  R}_p$. In this form it becomes clear that the times scales for the
modes, given by $(\zeta^l_p)^{-1}$, vary widely with the mode index
$p$, ranging from large for small $p$ to small for $p$ of the
order $N$. In the next sections, by separating the time-scales in this
manner, that the important physics is contained in the low modes and
that treating them correctly opens up a window of opportunity to take
large time steps in the numerical integration of Eq. (\ref{c3}).

Having said the above, we also note that the choice of the Rouse modes
in representing the dynamical equation is by no means unique. An
equivalent representation in terms of the polymer's fluctuation modes,
well-elaborated in one of our own publications \cite{leeuwen1} is as
follows. In terms of the bead positions ${\bf r}_n$ of the chain one
can expand the Hamiltonian around its ground state, which has a
configuration of a straight rod. The second term in this expansion,
involving the Hessian $\partial^2 {\cal H}/\partial {\bf
  r}_m\partial{\bf r}_n$, is also quadratic in the bead positions, but
it includes not only ${\cal H}^*$, but also some contribution from
$L_c$. Indeed, as shown in Ref. \cite{leeuwen1}, the corresponding
modes then yield the well-known transverse (bending) and longitudinal
(stretching) modes of a semiflexible chain, with eigenvalues
$\zeta^t_p$ and $\zeta^l_p$ respectively. [Of these, the longitudinal
modes are identical to the Rouse modes (\ref{c8}-\ref{c9}), which
explains our choice of notation for the eigenvalue in Eq. (\ref{c9}).]
Thus, an equivalent, and perhaps more natural, choice of representing
the dynamical equation (\ref{c3}) would be to use the longitudinal and
the transverse modes. Our experience, however, is that using the Rouse
mode representation makes the code faster and more robust for
parameters $T^*$ and $\nu$ typical for dsDNA, to which we stick to in
the rest of this paper (and also in our earlier publication
\cite{leeuwen1}).

\section{Time-integrated algorithm for the equation of motion for the
  polymer in mode representation\label{sec4}}

We start with the (obvious) statement that without the coupling term
${\bf H}_p$, the integration of the equations (\ref{c11}) is
straightforward. Each mode develops as an Ornstein-Uhlenbeck process,
which admits an exact solution. As this is the basis of our
refinements of the algorithm, we illustrate our method of
time-integration of the equation of motion for the polymer by
considering one scalar mode $R(t)$ with decay coefficient $\zeta$, a
coupling force $H(t)$ and random force $G(t)$. It is useful to first
make the substitution (c.f. the interaction representation in quantum
mechanics)
\begin{eqnarray} R(t) = \exp(-\zeta t)\, \tilde{R}(t),
\label{d1}
\end{eqnarray} leading to the equation for $\tilde{R} (t)$
\begin{eqnarray} \frac{d \tilde{R} (t)}{d t} =[H (t) + G (t)] \,
\exp(\zeta t).
\label{d2}
\end{eqnarray} Integrating this equation over a finite time interval
$\Delta t$ and multiplying the result with $\exp(- \zeta \Delta t)$
then yields
\begin{eqnarray} R (t + \Delta t) = \exp(- \zeta \Delta t)\, R(t) +
\overline{H} (t) + \overline{G} (t),
\label{d3}
\end{eqnarray} where $\overline{G} $ is given by
\begin{eqnarray} \overline{G} (t) = \int^{\Delta t}_0 dt' \,
\exp[\zeta(t' - \Delta t)] \, G(t +t'),
\label{d4}
\end{eqnarray} and likewise, $\overline{H} $ is given by
\begin{eqnarray} \overline{H} (t) = \int^{\Delta t}_0 dt' \,
\exp[\zeta(t' - \Delta t)] \, H(t +t').
\label{d5}
\end{eqnarray} The distribution of $\overline{G}(t)$ is, as an
integral (sum) over independent Gaussian random variables, i.e., a
Gaussian random variable with variance
\begin{eqnarray} w^2(\Delta t) = T^* [1 - \exp(-2 \zeta \Delta t)]
/\zeta;
\label{d6}
\end{eqnarray} i.e., in formula (\ref{d4}) the distribution reads
\begin{eqnarray} P( \overline{G} ) = \frac{1}{\sqrt{\pi} w(\Delta t)}
\exp \left( -\frac{\overline{G}^2}{2 w^2(\Delta t)} \right).
\label{d7}
\end{eqnarray} Note here that Equation (\ref{d3}) is an exact
substitute for the Langevin equation with an arbitrary time step.

From the above one sees that the use of the polymer's fluctuation
modes to time-integrate the equation of motion has two aspects:
\begin{itemize}
\item[(i)] If we manage to make $H$ small, we may treat the modes to
be evolving independently, with only a small perturbation due to the
coupling.
\item[(ii)] We need to find an expression for the integral
$\overline{H} (t)$, while we only have an expression for the initial
value $H(t)$.
\end{itemize} Clearly, the more successful we are with point (i), the
less severe point (ii) becomes.

\subsection{A more functional form of ${\bf H}_p$ for polymer dynamics\label{sec4a}}

The expression for ${\bf H_p}$ follows from Eqs. (\ref{c6}) and (\ref{c9})
\begin{eqnarray} 
{\bf H}_p = \frac{\partial L_c}{\partial
{\bf R}_p} = \sum_n \, \frac{\partial L_c} {\partial {\bf r}_n}
\phi_{n,p} = \sum_n \, [\hat{\bf u}_n - \hat{\bf u}_{n+1}] \,
\phi_{n,p},
\label{e1} 
\end{eqnarray} 
where $\hat{\bf u}_n={\bf u}_n/u_n$, is the unit bond vector, and
$u_n=|{\bf u}_n|$. By rearranging the summation variable $n$ we write
\begin{eqnarray}  
{\bf H}_p =\sum^N_{n=1} \hat{\bf u}_n \,
\chi_{n,p},
\label{e2}
\end{eqnarray} 
with
\begin{eqnarray} 
  \chi_{n,p}=\phi_{n,p} - \phi_{n-1,p}=
  2 \left(\frac{2}{N+1} \right)^{1/2} \sin \left(\frac{p\pi}{2(N+1)}
  \right) \, \sin \left(\frac{pn\pi}{N+1} \right).\nonumber\\
\label{e3} 
\end{eqnarray} 
Using the bond-length factor $b$ introduced in (\ref{a3}),
it is natural to write
\begin{eqnarray}
  \hat{\bf u}_n={\bf u}_n/b+\Delta\hat{\bf u}_n=(1-2\nu) {\bf u}_n+\Delta\hat{\bf u}_n.
\label{e4}
\end{eqnarray}
Inherent to Eq. (\ref{e4}) is the build-up of the following
approximation scheme, as we demonstrate below. In the limit of small $
T^*$ (i.e., high $\lambda$) --- e.g., $T^*=0.034$ for dsDNA --- the
chain does not stretch much, hence we expect $\Delta\hat{\bf u}_n$ to
be much smaller than ${\bf u}_n/b$; in other words, setting
$\Delta\hat{\bf u}_n$ to zero provides a rather good approximation for
$\hat{\bf u}_n$. Further, since ${\bf u}_n$ can be expressed as
\begin{eqnarray} {\bf u}_n = \sum^N_{q=1} \chi_{n,q} {\bf R}_q,
\label{e5} 
\end{eqnarray} 
with 
\begin{eqnarray} 
\sum^N_{n=1} \chi_{n,p} \, \chi_{n,q} =
\left[2 \sin \left(\frac{p\pi}{2(N+1)} \right) \right]^2
\delta_{p,q},
\label{e6} 
\end{eqnarray}
we can write
\begin{eqnarray} {\bf H}_p&=&(1 - 2 \nu) \left[2 \sin
      \left(\frac{p\pi}{2(N+1)} \right) \right]^2 \, {\bf R}_p + \Delta {\bf H}_p,
\label{e7} 
\end{eqnarray} 
where $\Delta {\bf H}_p$ is simply given by
\begin{eqnarray} \Delta {\bf H}_p = \sum^N_{n=1}
\chi_{n,p} \, {\bf u}_n \, [1/u_n - (1 - 2 \nu)].
\label{e11} 
\end{eqnarray} 
The first term of (\ref{e7}) can be combined with $- \zeta^l_p {\bf R}_p$ in  
Eq. (\ref{c11}), leading to the combination
\begin{eqnarray} 
\zeta_p=\zeta_p^l-(1 - 2 \nu) \left[2 \sin \left(\frac{p\pi}{2(N+1)}
  \right) \right]^2,
\label{e9}
\end{eqnarray} 
which curiously enough  is a reasonably good approximation of the
eigenvalue $\zeta^t_p$ for the $p$-th transverse mode \cite{leeuwen1}.
This allows us to rewrite Eq. (\ref{c11}) as
\begin{eqnarray} \frac{d {\bf R}_p (t)}{d \tau } =
  -\zeta_p{\bf R}_p+\Delta{\bf H}_p + {\bf G}_p,
\label{e10}
\end{eqnarray} 
with the hope that $\Delta{\bf H}_p$ remains small in comparison to
the full term ${\bf H}_p$. We will test this in Sec. \ref{sec4c}.

We note that the dynamical equation  (\ref{e10}) is still an exact representation 
of the Langevin equation (\ref{b1}). 
 
\subsection{Time-integration of $\Delta{\bf H}_p$ \label{sec4c}}

Following the notation of Eq. (\ref{d5}) we now discuss an
approximation for 
\begin{eqnarray}
\overline{\Delta {\bf H}_p (\tau)}=\int^{\Delta t}_0 dt' \,
\exp[\zeta^t_p(t' - \Delta t)] \,\Delta{\bf H}_p(t +t').
\label{f0}
\end{eqnarray}
In any time-forward integration process we clearly know the initial
value of the integrand in (\ref{f0}). We assume that the
integrand will decay in the interval $\Delta\tau$ with an exponent
comparable to the decay of the modes around $p$, as the strongest
correlation exists between nearby modes \cite{leeuwen1}. A further
assumption we make here is that since $\Delta{\bf H}_p$ contains
purely the bond-length fluctuations, which are part of the
longitudinal fluctuations of the chain, we expect the exponent to be
equal to $\zeta^l_p$; leading us to the approximation
\begin{eqnarray} \Delta {\bf H}_p (\tau+\tau') \simeq
\Delta {\bf H}_p (\tau) \exp (-\zeta^l_p \tau').
\label{f1} 
\end{eqnarray} 
Then the integral (\ref{d5}) simply reduces to
\begin{eqnarray} 
  \overline{\Delta {\bf H}_p} (\tau) \simeq
  \Delta {\bf H}_p (\tau) \frac{\exp(-\zeta_p \Delta \tau) - \exp
    (-\zeta^l_p \Delta \tau) }{\zeta^l_p-\zeta_p}.
\label{f2} 
\end{eqnarray} 
For modes where $\zeta_p \Delta\tau$ and $\zeta^l_p \Delta\tau$ are
both small, the expression reduces to
\begin{eqnarray} 
\overline{\Delta H_p (\tau)} \simeq \Delta
H_p (\tau) \, \Delta \tau, \quad \quad \quad p \quad {\rm small}.
\label{f3} 
\end{eqnarray} 
Indeed, this is precisely what one would expect for the modes that do
not decay in the interval $\Delta \tau$. Similarly, for the modes
where $\zeta^l_p \Delta \tau$ is large one gets
\begin{eqnarray} \overline{\Delta {\bf H}_p} (\tau) \simeq
\Delta {\bf H}_p (\tau) \frac{\exp(-\zeta_p \Delta
\tau)}{\zeta^l_p-\zeta_p},
\label{f4} 
\end{eqnarray} 
i.e.  an exponentially small contribution. In other words, the form
(\ref{f1}) gives a smooth suppression of the coupling between the
high-$p$ modes, depending on the choice of $\Delta \tau$. For time
steps $\Delta \tau$ in which the high modes are equilibrated, we treat
them as independent modes. In the extreme limit of very large $\Delta
\tau$ all the modes become independent. That limit clearly misses the
important non-linear effects between the modes, which essentially puts
a limit on how large $\Delta\tau$ we can get away with.

\section{Testing the time-integration algorithm for
  $\mbox{ds}$DNA \label{sec5}}

Having explained the time-integration algorithm in the previous
section in general terms within this bead-spring model, we now set out
to test it on a single dsDNA chain. However, before we do so, it is 
imperative to us that we check whether Eq. (\ref{e10}) provides a 
reasonable time-integration scheme. That starts with a comparison of 
the ${\bf H}_p$ and $\Delta {\bf H}_p$ terms in Eq. (\ref{e10}).
\begin{figure*}
\begin{center}
\includegraphics[width=0.6\linewidth]{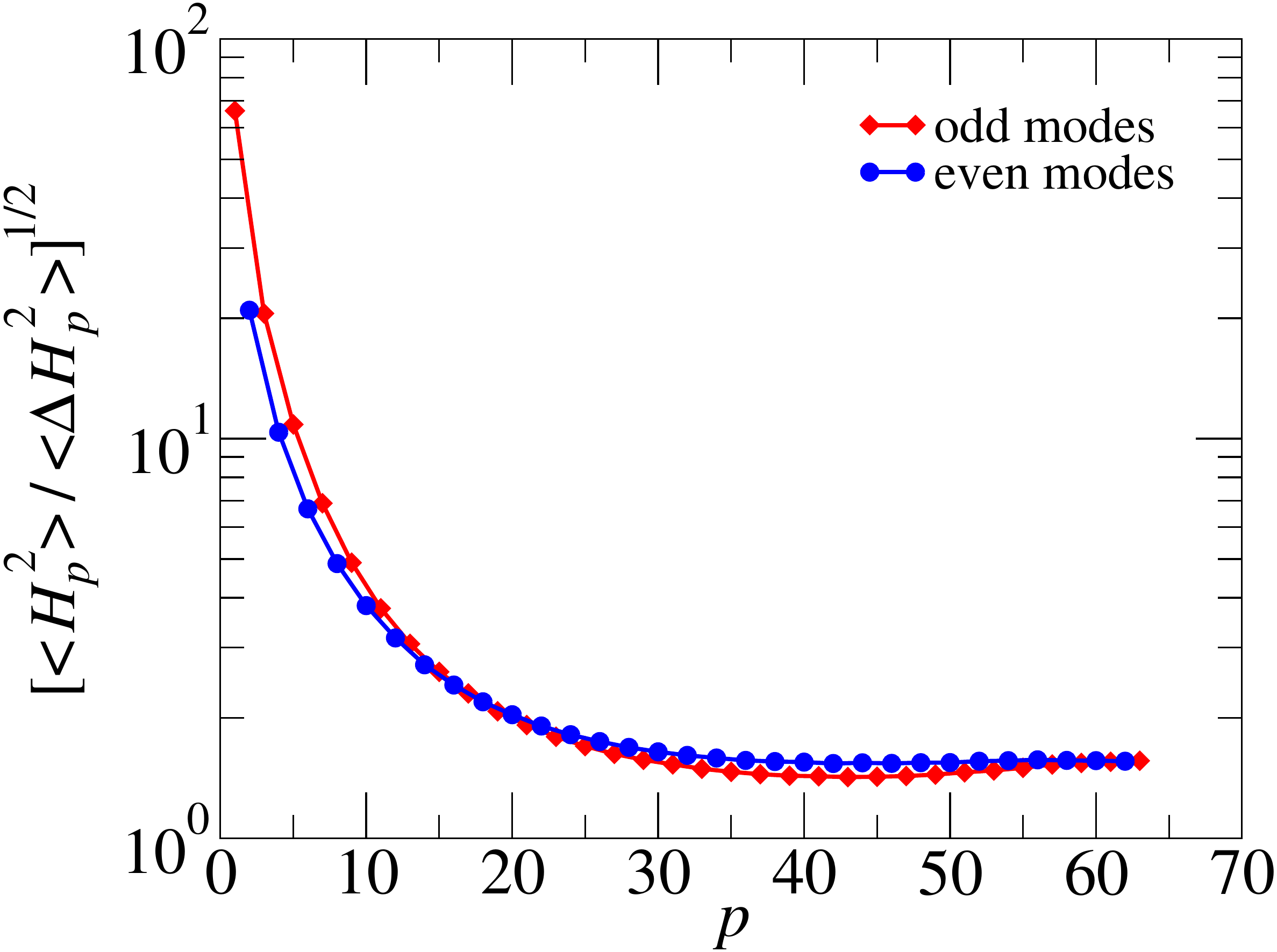}
\end{center}
\caption{(Color online) The ratio $[\langle H_p^2\rangle/\langle\Delta
  H_p^2\rangle]^{1/2}$ for a dsDNA chain of length
  $N=63$.\label{fig1}}
\end{figure*}

\subsection{Comparing ${\bf H}_p$ and $\Delta {\bf H}_p$ for dsDNA\label{sec5a}}

Both ${\bf H}_p$ and $\Delta {\bf H}_p$ are fluctuating quantities, so
a proper comparison between them would be to plot the ratio of
$[\langle H_p^2\rangle/\langle\Delta H_p^2\rangle]^{1/2}$ as a
function of $p$. In doing so we note that odd-$p$ modes are even under
reversal of renumbering the beads from $n$ to $N-n$ and even-$p$ modes
are odd under this reversal. Since the ground state is even under this
reversal, it means that the odd-$p$ modes are excited more by thermal
fluctuations. We therefore plot this ratio separately for the even and
the odd modes in Fig. \ref{fig1} for a dsDNA with $N=63$, i.e., a
dsDNA chain 64 basepairs long. The plot shows the approximation scheme
(\ref{e7}) in action --- for low $p$-values, i.e., modes corresponding
to large length-scales, the remainder $\Delta{\bf H}_p$ is only a
fraction of ${\bf H}^{(0)}_p$. This opens up a systematic way of
dealing with $\Delta{\bf H}_p$ that still couples the different
(Rouse) modes, which we exploit in the next subsection.

\subsection{Testing the vulnerability of the algorithm to enlarging
  $\Delta\tau$ for dsDNA\label{sec5b}}

As already pointed in Sec. \ref{sec4a}, with the approximations
(\ref{f1}-\ref{f4}) we cannot limitlessly increase $\Delta\tau$. We
now test numerically on dsDNA how far we can go on with increasing
$\Delta\tau$ for $N=63, 127, 255$ and $511$. In other words, we obtain
the values of $\Delta\tau_{\text{max}}$ for these values of $N$. The
quantities we track, collectively denoted by $Q(t)$, in order to
determine $\Delta\tau_{\text{max}}$ are the autocorrelation functions
in time of (i) the end-to-end vector and (ii) the middle bond, and
(iii) the mean-square displacement (msd) of the middle bead wrt the
position of the centre-of-mass of the chain. Our test procedures are
divided into two groups: the equilibrium values for these quantities,
collectively denoted as $Q(0)$ and their dynamical behaviour. The test
procedure is as follows.
\begin{figure*}
\begin{center}
\includegraphics[width=0.48\linewidth]{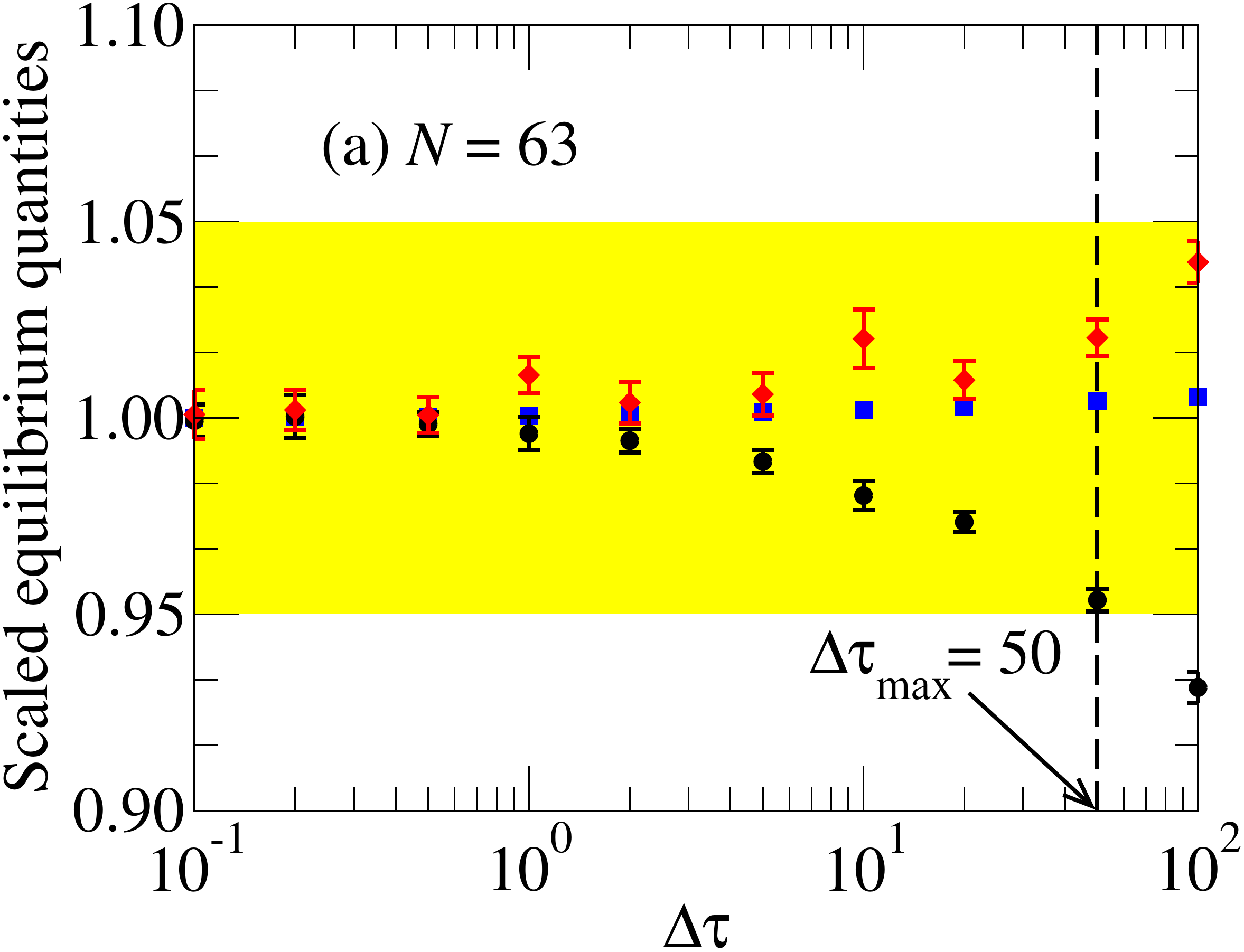}
\hspace{5mm}\includegraphics[width=0.48\linewidth]{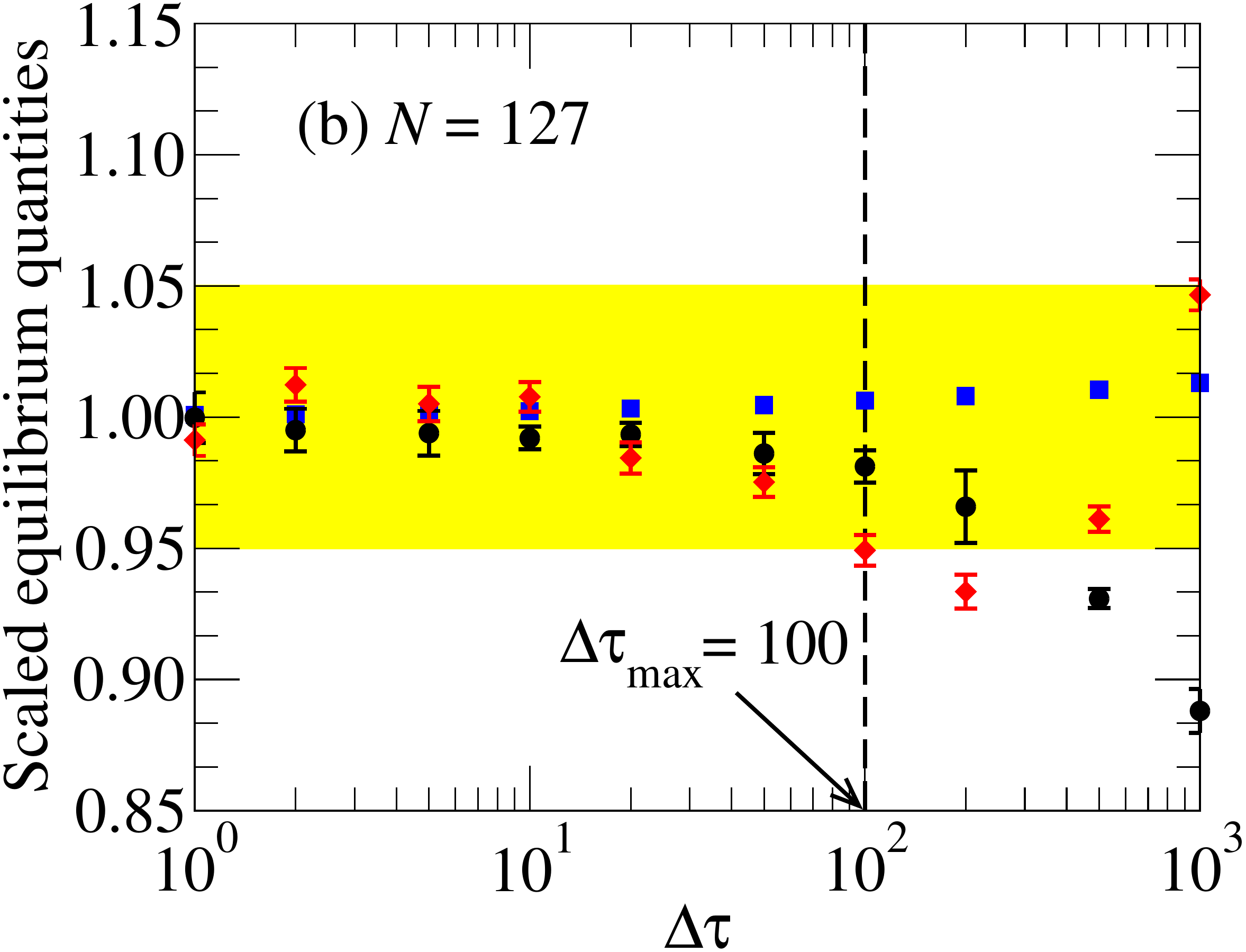}
\includegraphics[width=0.48\linewidth]{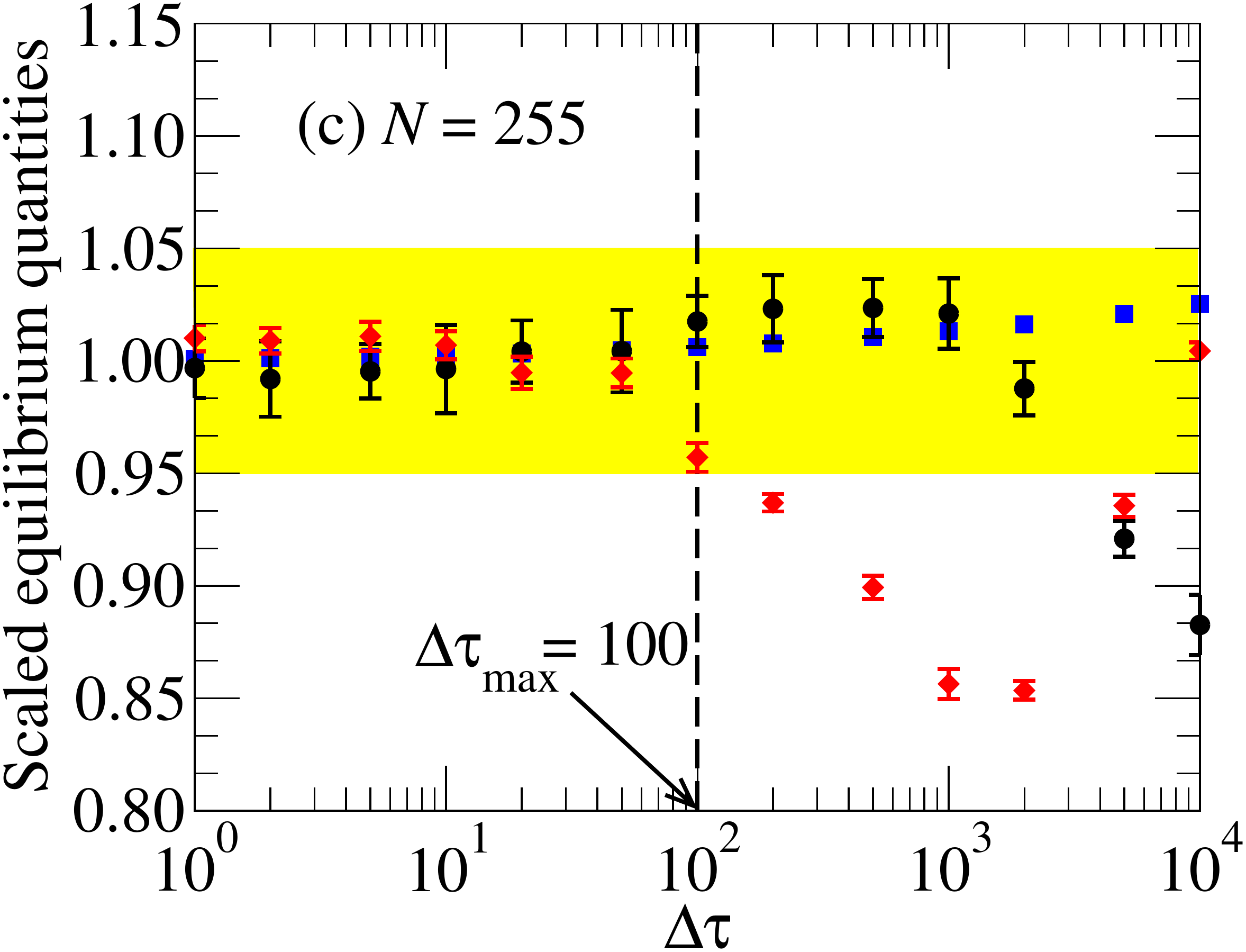}
\hspace{5mm}\includegraphics[width=0.48\linewidth]{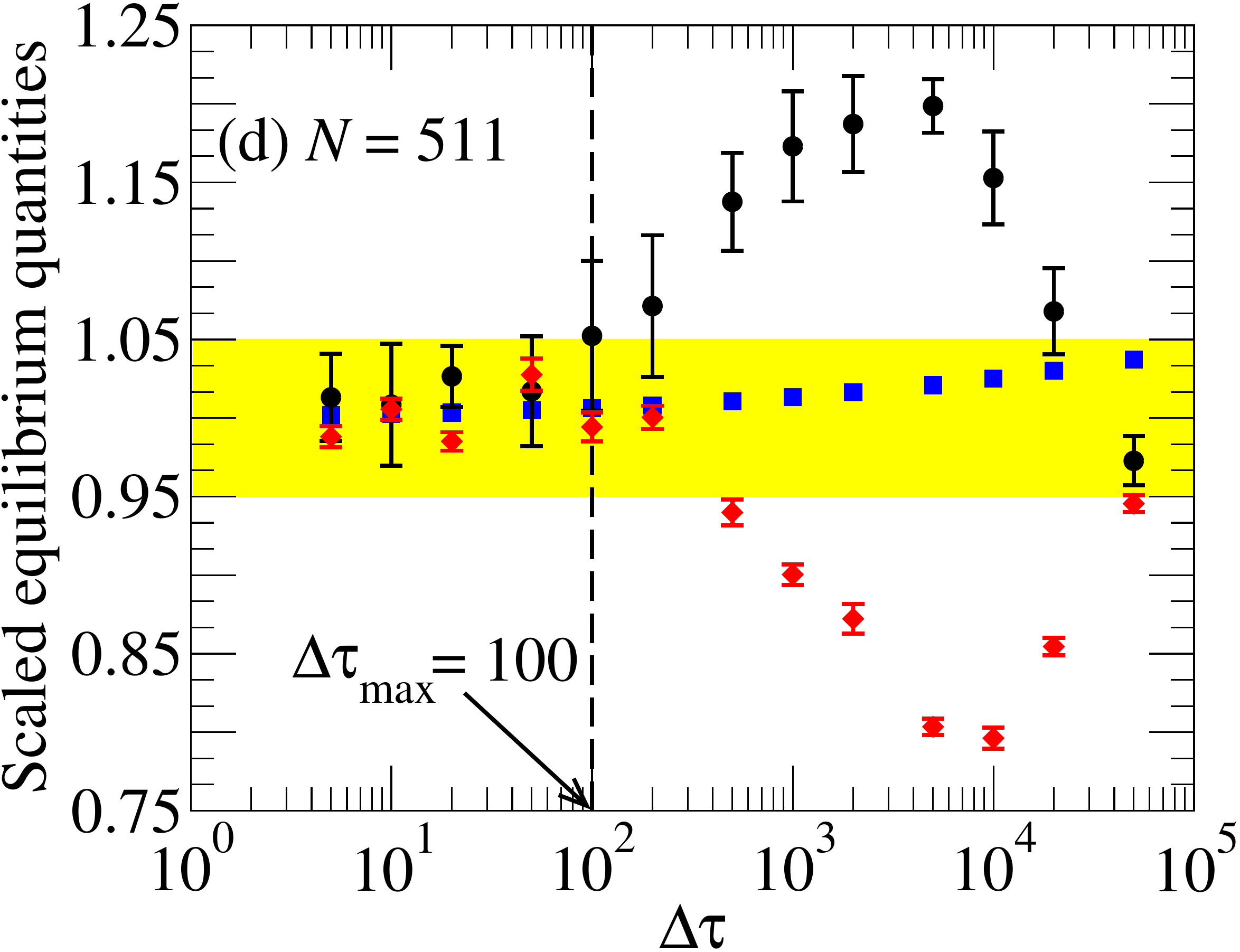}
\end{center}
\caption{(Color online) Determination of $\Delta\tau_{\text{max}}$ from the
  equilibrium values $Q(0)$. The autocorrelation functions in time of
  the end-to-end vector are shown in black circles, the middle bond in
  blue squares, and the mean-square displacement (msd) of the middle
  bead wrt the position of the centre-of-mass of the chain in red
  diamonds. The data are rescaled by their MC values: (a) $N=63$, (b)
  $N=127$, (c) $N=255$ and (d) $N=511$. The error bars for the
  autocorrelation functions in time of the middle bond are not shown
  since they are smaller than the symbol size. The yellow band
  represents $5\%$ validity thresholds. See text for
  details.\label{fig2}}
\end{figure*}

In the first group, we determine the quantities (i-iii) for several
values of $\Delta\tau$, namely (a) $N=63: \Delta\tau=$0.1, 0.2, 0.5,
1, 2, 5, 10, 20, 50 and 100, (b) $N=127: \Delta\tau=$1, 2, 5, 10, 20,
50, 100, 200, 500 and 1000, (c) $N=255: \Delta\tau=$1, 2, 5, 10, 20,
50, 100, 200, 500, 1000, 2000, 5000 and 10000, (d) $N=511:
\Delta\tau=$ 5, 10, 20, 50, 100, 200, 500, 1000, 2000, 5000, 10000,
20000 and 50000. The data are averaged over 13 realisations of run
length $\tau=8\times10^7$ for $N=63$, $\tau=8\times10^8$ million for
$N=127$, $\tau=8\times10^9$ for $N=255$ and $\tau=4\times10^{10}$ for
$N=511$ for each realisation. In the first group of tests we determine
their equilibrium values as a function of $\Delta\tau$. As benchmarks
of these equilibrium quantities we also perform Monte Carlo (MC)
simulations, which are carried out again by using the mode
representation, permitting one to take large MC steps in the slow
modes and small steps in the fast modes. We accept the runs as valid
if all measured observables deviate at most $5\%$ from their MC
values, the other ones we reject as invalid, leading to a set of
$\Delta\tau_{\text{max}}$ values for each $N$.
\begin{figure*}
\begin{center}
\includegraphics[width=0.46\linewidth]{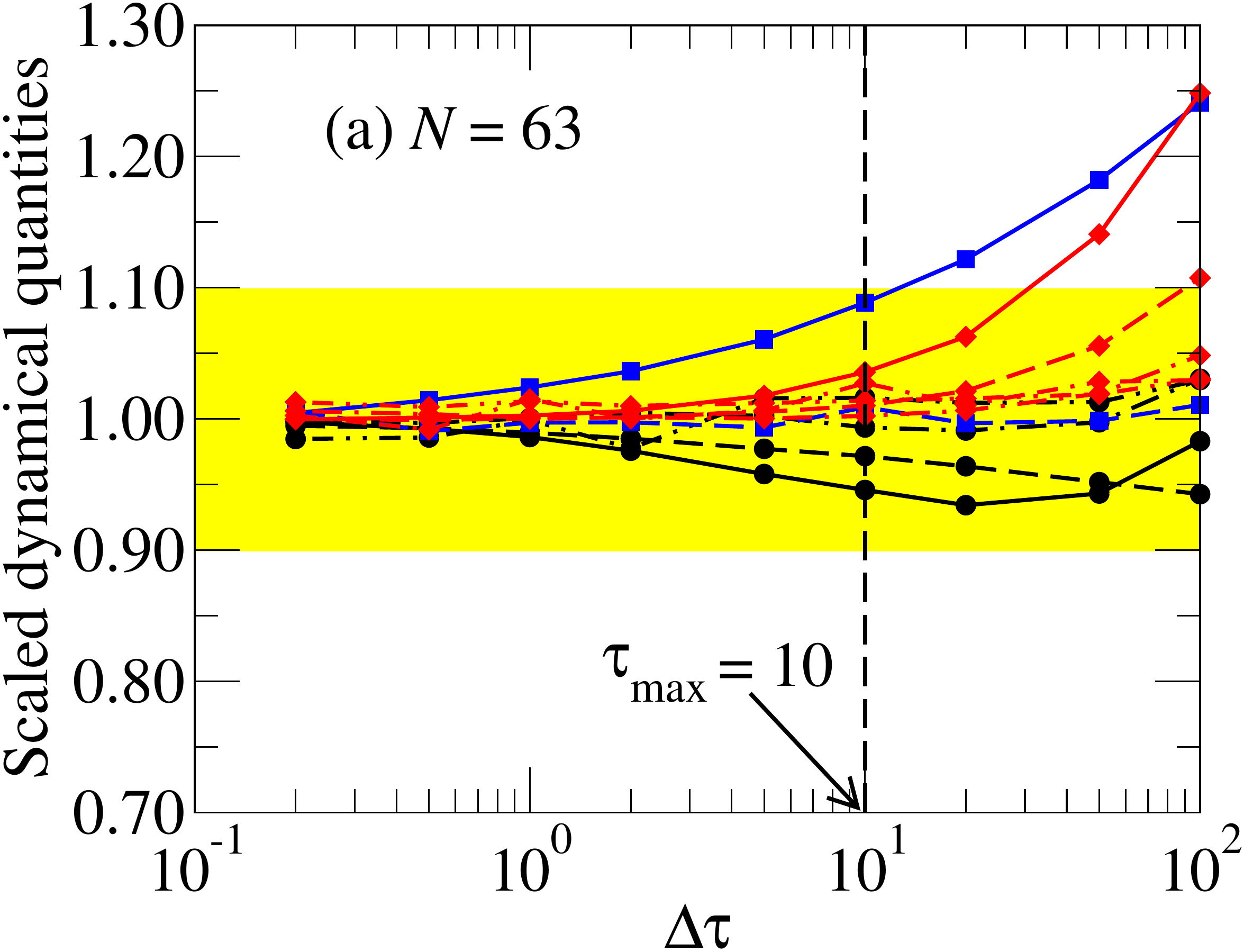}
\hspace{5mm}\includegraphics[width=0.46\linewidth]{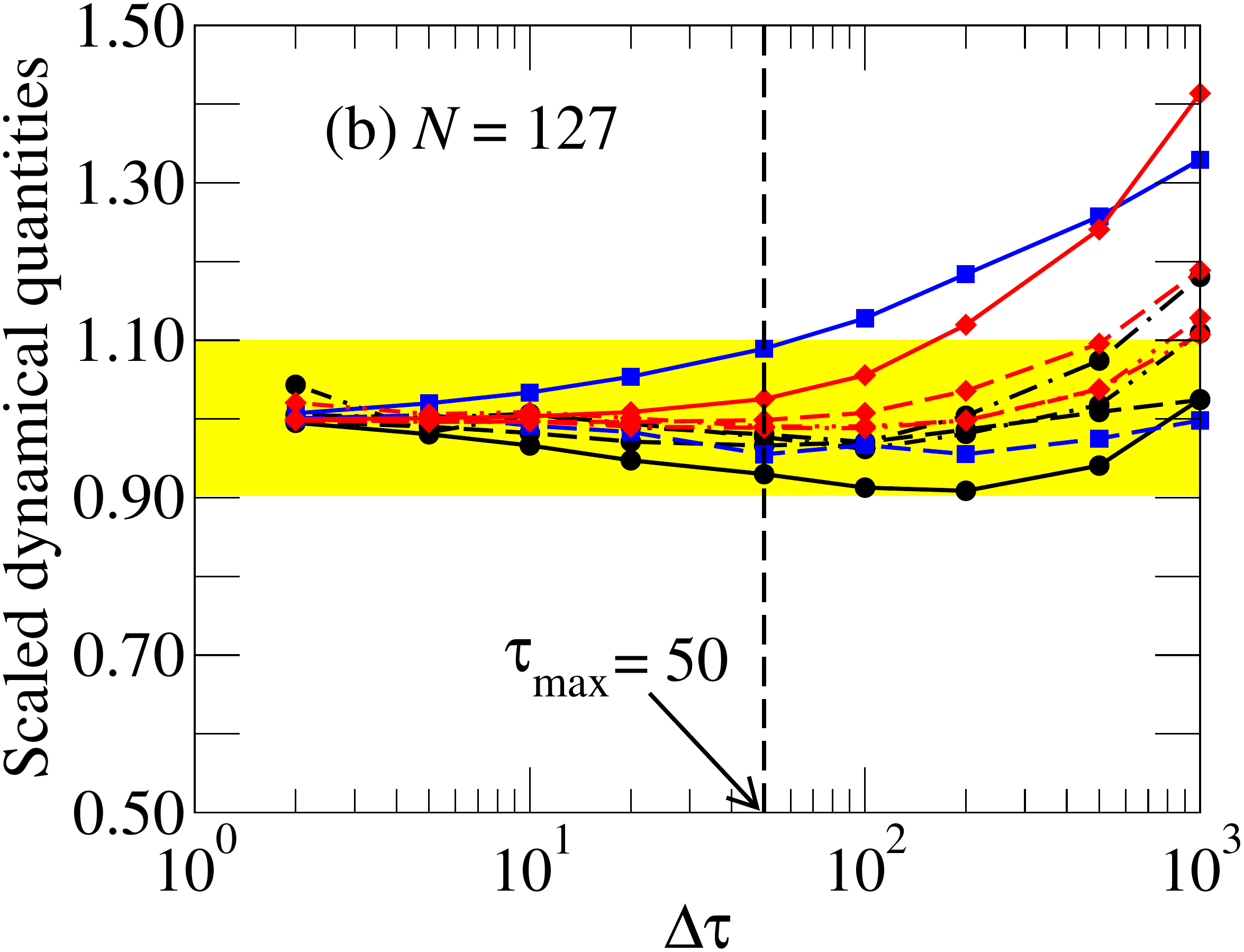}
\includegraphics[width=0.46\linewidth]{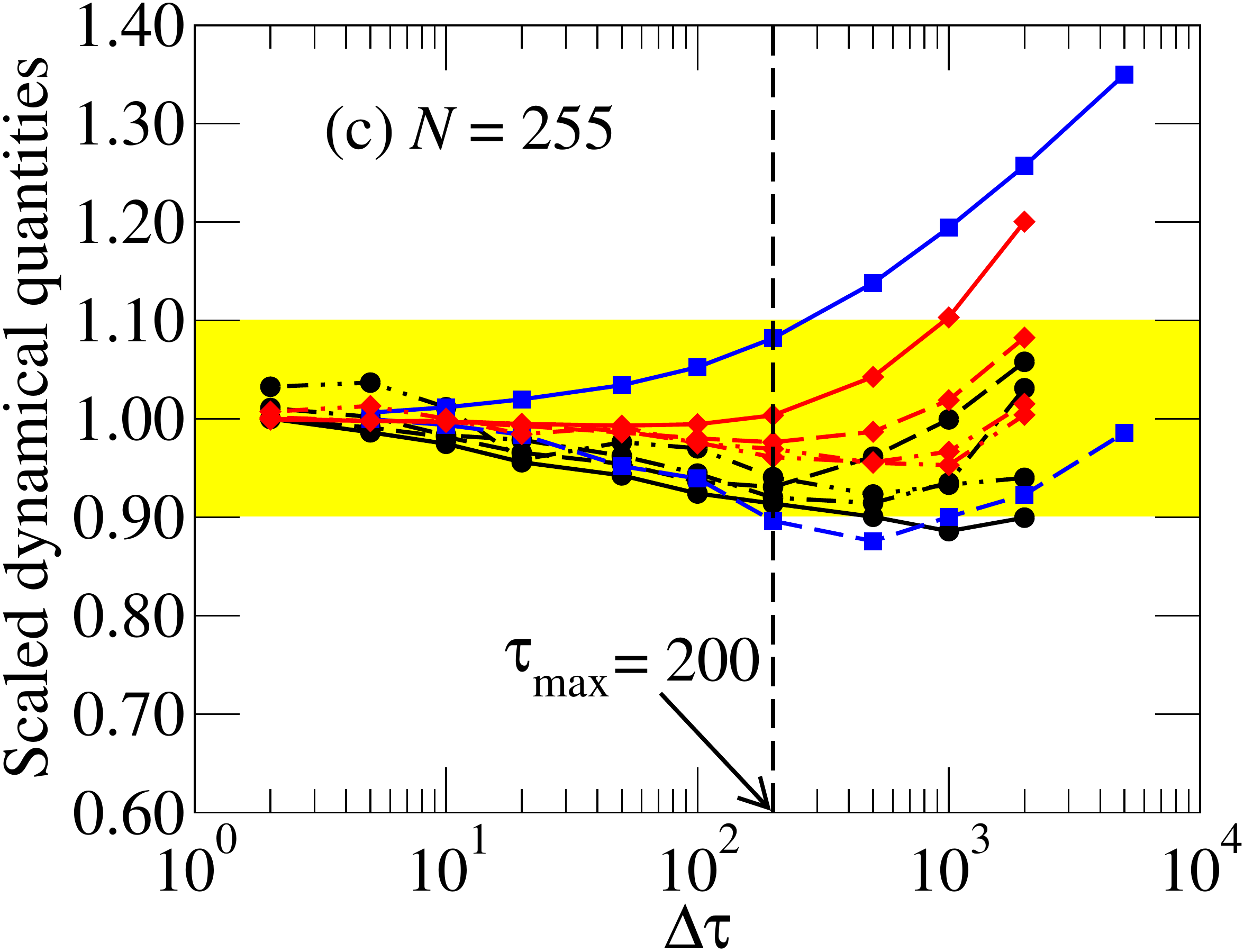}
\hspace{5mm}\includegraphics[width=0.46\linewidth]{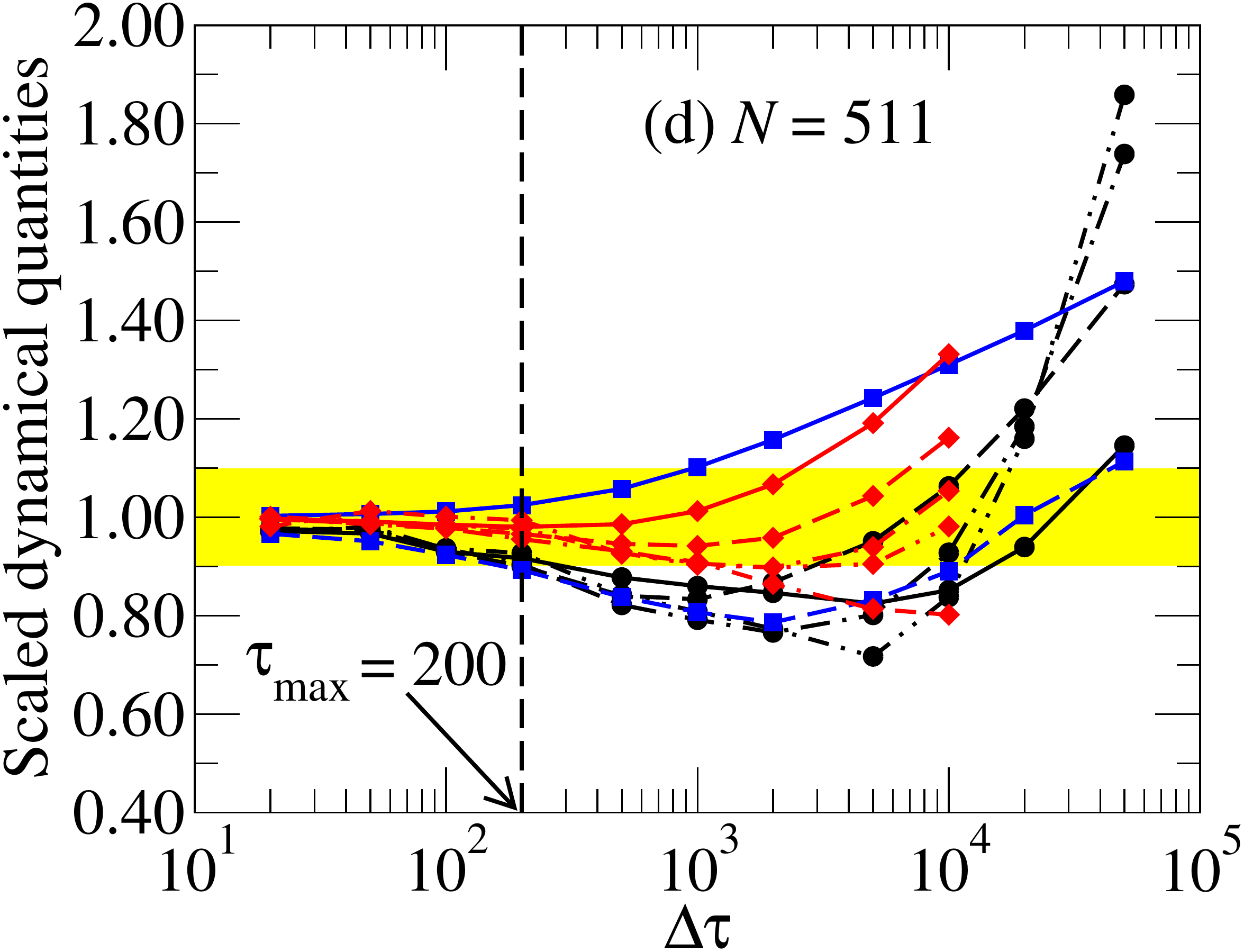}
\end{center}
\caption{(Color online) Determination of $\Delta\tau_{\text{max}}$ from 
  $Q(t)$. The autocorrelation functions in time of the end-to-end vector 
  are shown in black circles, the middle bond in blue squares, and the
  mean-square displacement (msd) of the middle bead wrt the position
  of the centre-of-mass of the chain in red diamonds. The data are
  rescaled by $\tilde Q(\Delta\tau_{\text{min}})$. (a) $N=63$
  ($\Delta\tau_{\text{min}}=0.1$): $\tau=100$ (solid line), 1000
  (long-dashed), 10000 (dashed-dot-dashed), 100000
  (dashed-dot-dot-dashed); (b) $N=127$ ,
  ($\Delta\tau_{\text{min}}=1$): $\tau=1000$ (solid line), 10000
  (long-dashed), 100000 (dashed-dot-dashed), 1000000
  (dashed-dot-dot-dashed); (c) $N=255$
  ($\Delta\tau_{\text{min}}=1$): $\tau=10000$ (solid line), 100000
  (long-dashed), 1000000 (dashed-dot-dashed), 10000000
  (dashed-dot-dot-dashed) and (d) $N=511$
  ($\Delta\tau_{\text{min}}=5$): $\tau=100000$ (solid line), 1000000
  (long-dashed), 10000000 (dashed-dot-dashed), 100000000
  (dashed-dot-dot-dashed). The yellow band represents $10\%$
  validity thresholds. See text for details.\label{fig3}}
\end{figure*}

The procedure is demonstrated in Fig. \ref{fig2}. We scale the
equilibrium values by the corresponding MC ones, which means that the
$y$-values of the rescaled equilibrium quantities should lie between
0.95 and 1.05, indicated by the yellow band representing our
acceptance threshold. The highest $\Delta\tau$ values, for which all
the equilibrium quantities --- taking into consideration the error
bars --- fall within the yellow band gets us the
$\Delta\tau_{\text{max}}$ values for each $N$ for the first group of
test.
\begin{figure*}
\begin{center}
\includegraphics[width=0.48\linewidth]{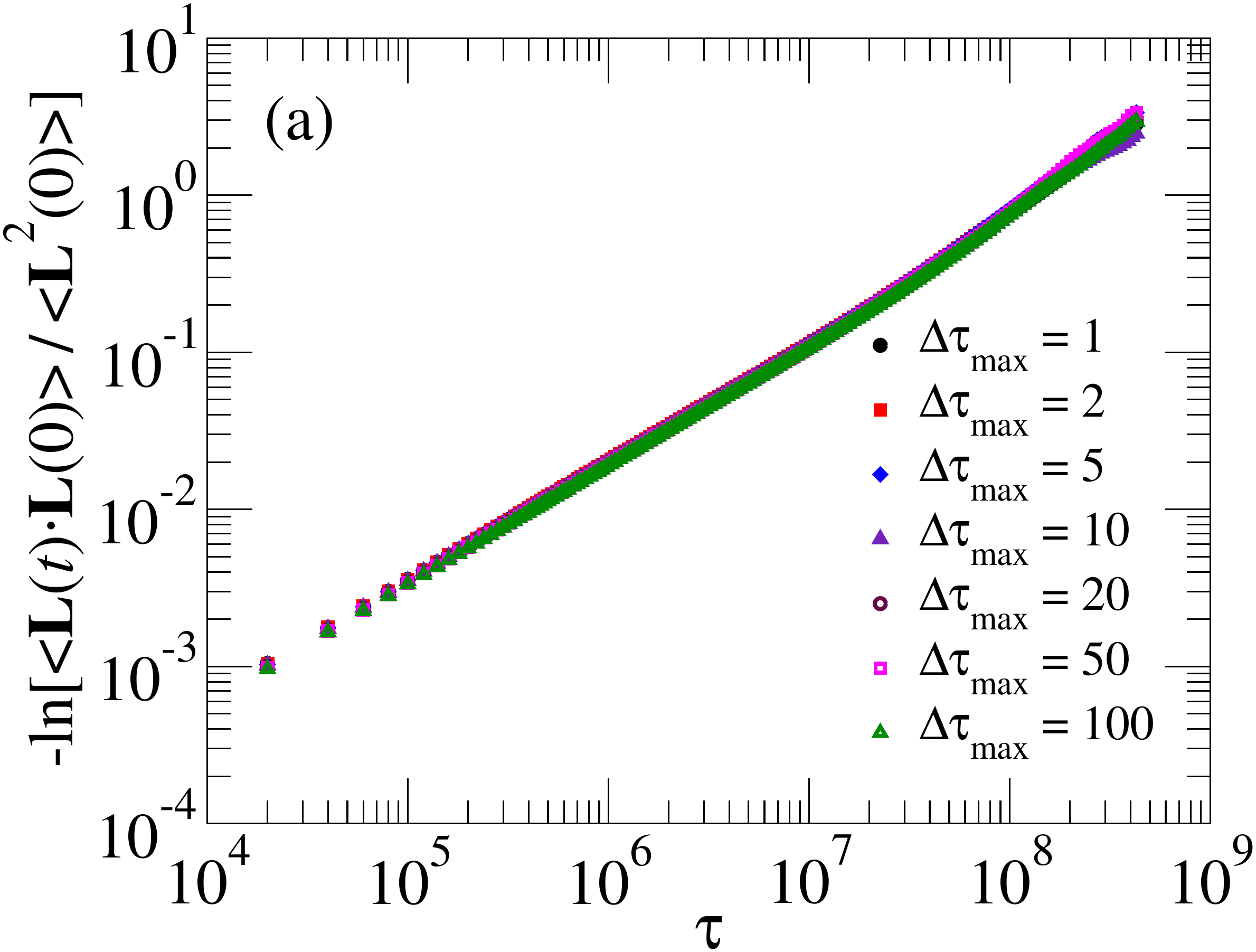}
\hspace{5mm}\includegraphics[width=0.48\linewidth]{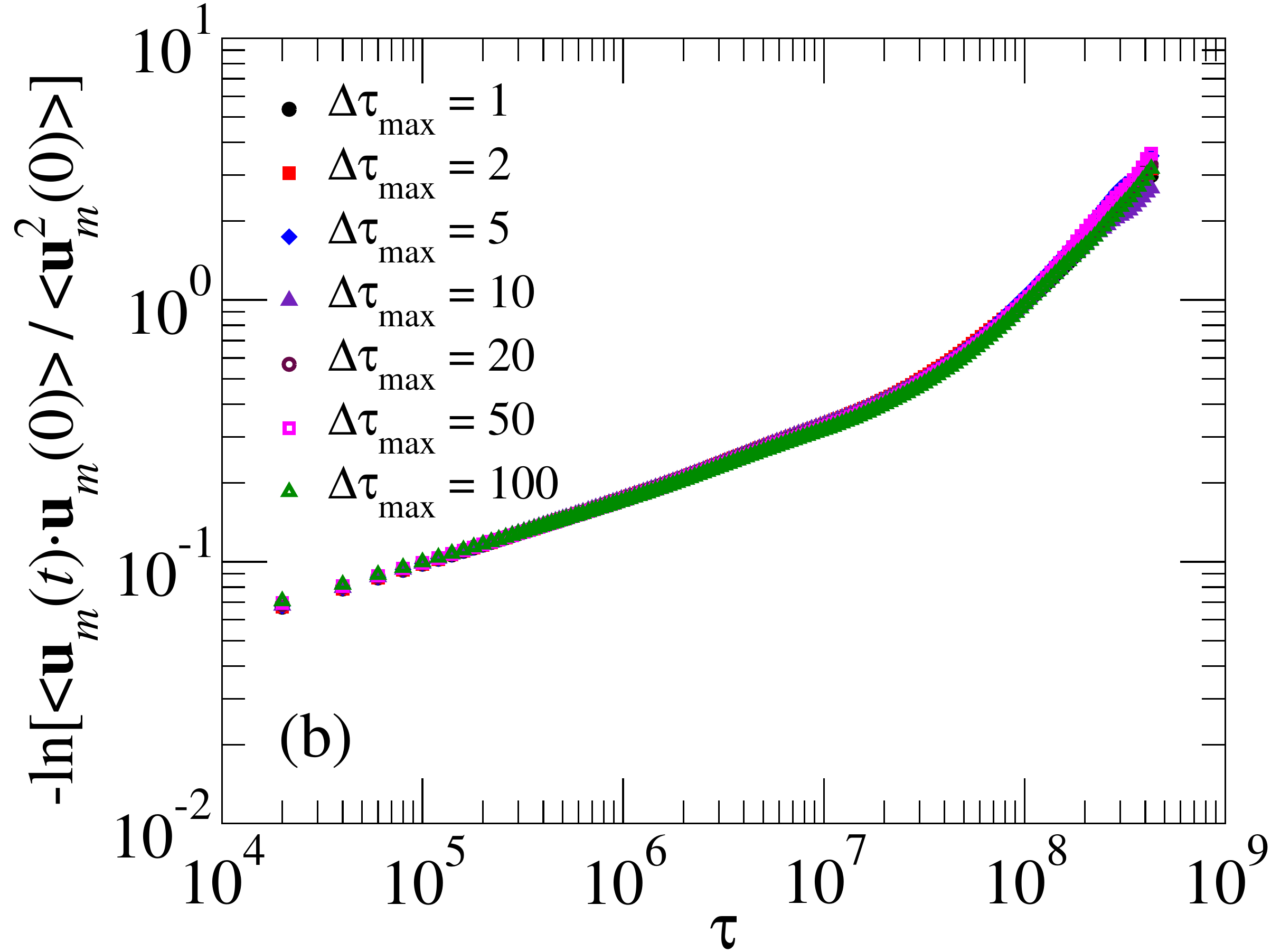}
\includegraphics[width=0.48\linewidth]{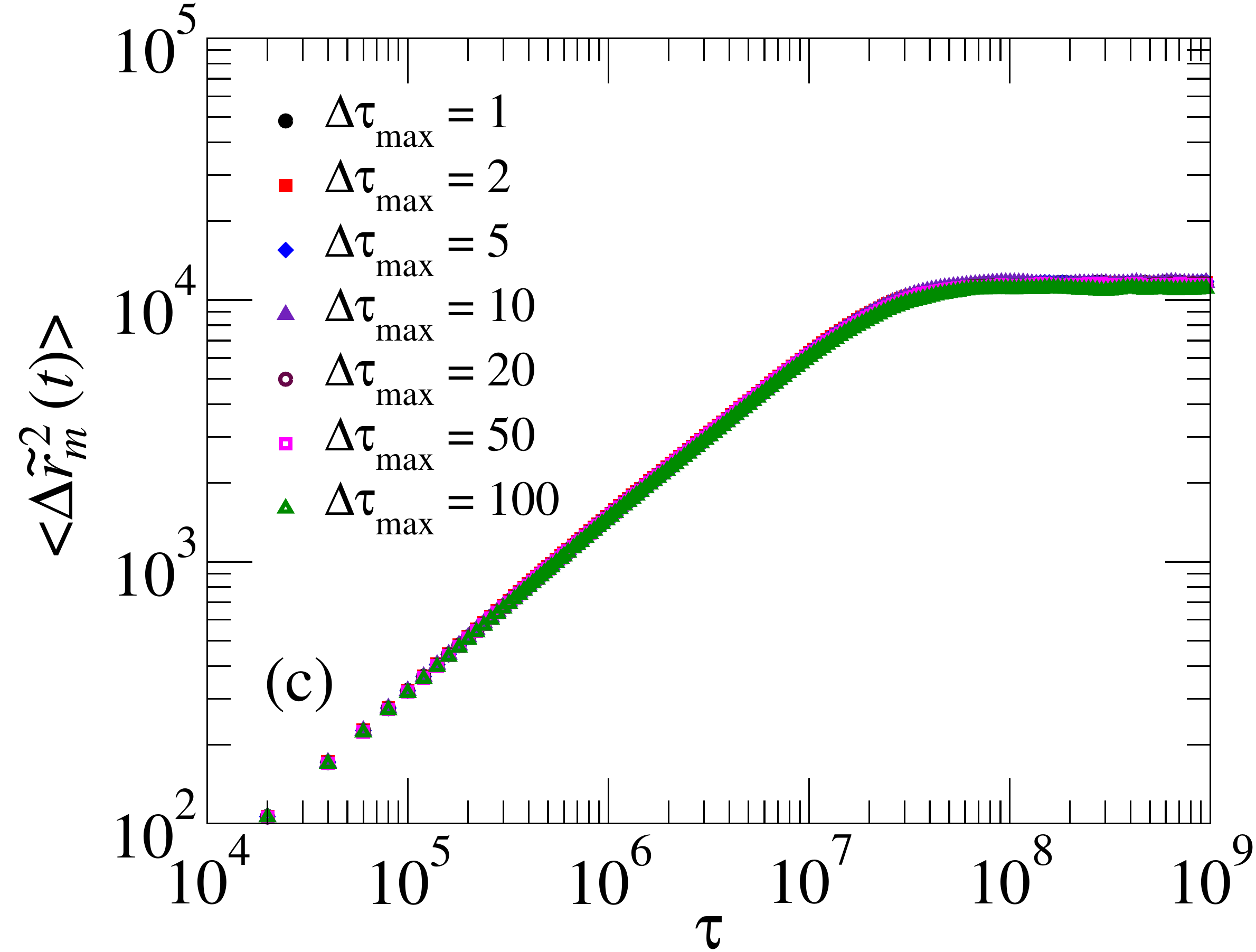}
\end{center}
\caption{(Color online) The $Q(t)$ curves for 
  $\Delta\tau=\Delta\tau_{\text{min}}$ to $\Delta\tau_{\text{max}}$ 
  for $N=255$: the normalised autocorrelation function of the end-to-end 
  vector ${\mathbf L}$ (a) and the middle bond vector ${\mathbf u}_m$ 
  (b), and the means-square displacement $\langle\Delta\tilde 
  r^2_m(t)\rangle$ of the middle bead measured wrt the centre-of-mass 
  of the chain (c). The data for all values of $\Delta\tau$ coincide, 
  as they should. The figure thus demonstrates the validity of our 
  procedure. \label{fig4}}
\end{figure*}

In the second group of tests we use time-dependent quantities
$Q(t)$. In this case MC simulations are of no help, so we treat the
lowest value of $\Delta\tau$ for each $N$ as the benchmark. Let us
describe the procedure for $N=63$, for which the lowest value of
$\Delta\tau$ equals $\Delta\tau_{\text{min}}=0.1$. We choose a few
fixed values of the time $\tau$, such as $\tau=100$, 1000, 10000 and
100000; first obtain the quantities $Q(\tau)$ and numerically
differentiated quantity $dQ(\tau)/d\tau$ for all values of
$\Delta\tau$, and thereafter the effective decay constant for
$Q(\tau)$, i.e., ratio $\tilde
Q(\tau)=\displaystyle{\frac{1}{Q(\tau)}\frac{dQ(\tau)}{d\tau}}$. [Clearly,
for a given value of $\tau$, $\tilde Q(\tau)$ is also a function of
$\Delta\tau$, i.e., $\tilde Q(\tau)\equiv\tilde Q(\tau,\Delta\tau)$.]
We then demand that at these values of $\tau$ the ratio $\tilde
Q(\Delta\tau)/\tilde Q(\Delta\tau_{\text{min}})$ does not deviate from
unity by more than $10\%$. The $10\%$ is chosen by the following
criterion: the statistical errors in the quantities $Q(t)$ are
typically of order $2\%$, which accumulate to $\sim4\%$ for $\tilde
Q(t)$, to which we need to add our $5\%$ criterion as explained above,
and further round their sum off to $10\%$. (The larger tolerance for
the dynamical variables is a consequence of a lack of clean benchmark
data, which was provided by MC simulations for equilibrium
observables.) The result of this procedure is presented in
Fig. \ref{fig3} --- note that this (numerical) procedure is prone to
noise more than it has been for the first group that involved $Q(0)$,
so we only carry out the procedure for which the procedure is not
spoiled by noise in the data.

The results from the two figures \ref{fig2}-\ref{fig3} are summarised
in Table \ref{table1}. The final value of $\Delta\tau_{\text{max}}$
for any given value of $N$ is clearly the smaller one emerging from
the two groups. Said differently, use of the final values of
$\Delta\tau_{\text{max}}$ (as it appears in Table \ref{table1}) in
simulations means that the data for $Q(t)$ should not differ from each
other by more than $5\%$ at any time. As an example of the validity of
our procedure, we plot the $Q(t)$ curves for $N=255$ in
Fig. \ref{fig4} --- for all $\Delta\tau$ values between
$\Delta\tau_{\text{min}}$ and $\Delta\tau_{\text{max}}$ the curves are
on top of each other as they should be.
\begin{table*}
\begin{center}
\begin{tabular}{c||c||c||c}
  \hline\hline$\quad N\quad$&$\quad\Delta\tau_{\text{max}}$ from equilibrium$\quad$ &$\quad\Delta\tau_{\text{max}}$
  from dynamical$\quad$ &$\quad$ final $\Delta\tau_{\text{max}}\quad$\tabularnewline
  &quantities $Q(0)$&quantities $Q(t)$& $\quad$min[columns 2 and 3]$\quad$\tabularnewline\hline\hline
  63&50&10&10\tabularnewline\hline
  127&100&50&50\tabularnewline\hline
  255&100&200&100\tabularnewline\hline
  511&100&200&100\tabularnewline\hline\hline
\end{tabular}
\end{center}
\caption{List of $\Delta\tau_{\text{max}}$ values for the values of $N$
  studied in this paper for dsDNA.\label{table1}}
\end{table*}

\section{Coarse-graining in our model\label{newsec}}

Up until now we have chosen the average spacing between the beads to
coincide with the length of a dsDNA basepair $\approx0.33$ nm. There
is nothing special about this choice. In this section we explore the
case when the average spacing between the beads is larger than the
length of a dsDNA basepair, i.e., coarse-graining in our model, and
its consequences.

While coarse-graining, we note that we have to consistently conform to
the force extension curve for the dsDNA. The force extension relation
proposed by Wang {\it et al.}  \cite{wang} has the form
\begin{equation} 
  \frac{F l_p}{k_B T} =\frac14 \left[ {1 -  \frac{\langle L \rangle}{L_c} + \frac{F}{K_0}}
  \right]^{-2} -  \frac14 +  \frac{\langle L \rangle}{L_c} - \frac{F}{K_0},
  \label{x1}
\end{equation} 
where $F$ is the applied force and $\langle L \rangle$ the average
extension. The equation contains two empirical parameters: the
persistence length $l_p$ and the force constant $K_0$. We convert them
in dimensionless quantities as 
\begin{equation} 
  r=\frac{l_p}{a}, \quad \quad \quad y=\frac{l_p  K_0}{k_B T},
  \label{x2}
\end{equation}  
where $a$ is the length of a basepair. The model parameters $T^*$ and
$\nu$ are calculated by a fit to this force-extension curve as
\cite{leeuwen}
\begin{eqnarray} 
  T^* = \frac{(2 r^2 + y) r}{y^2} \quad
  \quad \quad {\rm and} \quad \quad \quad \nu= \frac{r^2}{2 r^2 + y}.
  \label{x3} 
\end{eqnarray}

We now make the choice for the discretization distance to be a
multiple of the length of a basepair, and represent it by $ka$. With
this choice, the parameter $l_p/a$ for the force extension curve
changes from $r$ to $r_k=r/k$. The force constant $K_0$ and the
persistence length $l_p$ must remain the same for the coarse-grained
description of the chain, implying that $y$ remains the same as well.
The new parameters $T^*_k$ and $\nu_k$ for our model then read
\begin{equation} 
  T^*_k = \frac{(2 r^2 + k^2 y) r}{k^3 y^2}, \quad \quad \quad \nu_k = \frac{r^2}{2 r^2 + k^2 y}.
  \label{x4}
\end{equation}  
Thus with increasing $k$ the model travels through a sequence of
parameter points $(T^*_k, \nu_k)$ all leading to the same
force-extension curve (\ref{x1}).  As long as $k\ll r$ the loss of
information due to coarse-graining will be small and the parameters
$(T^*_k, \nu_k)$ will adequately describe the polymer at the chosen
coarse-grained level. Moreover, while coarse-graining we must remember
that the average inter-bead spacing should not exceed the persistence
length, i.e., $k$ should stay well below $r$. In Table \ref{table3}
below we give a set of parameters $(T^*_k,\nu_k)$ for a number of
values of $k$. 
\begin{table}[h]
\begin{center}
\begin{tabular}{c||c||c}
\hline \hline$k$ & $T^*_k$ & $\nu_k$\tabularnewline\hline\hline
$\quad1\quad$  &  $\quad0.034\quad$  &  $\quad0.35\quad$ \tabularnewline\hline
$\quad2\quad$  &  $\quad0.008\quad$  &  $\quad0.1875\quad$ \tabularnewline\hline
$\quad5\quad$  &  $\quad0.002192\quad$  &  $\quad0.0437956 \quad$ \tabularnewline\hline
$\quad10\quad$  &  $\quad0.001024\quad$  &  $\quad0.0117188\quad$ \tabularnewline\hline
$\quad12\quad$  &  $\quad0.000847\quad$  &  $\quad0.00819672\quad$ \tabularnewline\hline
$\quad15\quad$  &  $\quad0.000673\quad$  &  $\quad0.00527704 \quad$ \tabularnewline\hline
$\quad20\quad$  &  $\quad0.000503\quad$  &  $\quad0.00298211\quad$ \tabularnewline\hline
\end{tabular}
\end{center}
\caption{Parameter values for dsDNA for our model under 
  coarse-graining. With increasing $k$ the model travels through 
  a sequence of parameter points $(T^*_k, \nu_k)$, all leading to the 
  same  force-extension curve (\ref{x1}). The case $k=1$ corresponds
  to the situation when the inter-bead spacing is the length of a
  basepair, $\approx0.33$ nm. \label{table3}}
\end{table}

\subsection{The $\nu \rightarrow 0$ limit and the inextensible WLC
  model\label{limit}}

In Table \ref{table3} we observe that with increasing degree of
coarse-graining the value of $\nu$ becomes progressively
smaller. Since the condition $k<r$ provides an upper bound for $k$,
the range of $k$ is rather small for the dsDNA to get really close to
zero. (This is however not the case for f-actin, for which the table
analogous to Table \ref{table3} can be found in Appendix A.)  In the
limit $\nu=0$, i.e., $\lambda \rightarrow \infty$ at fixed $\kappa$
--- this is the same limit $k_x \rightarrow \infty$ for the extensible
WLC, c.f. Sec. \ref{newseca} --- our model physically approaches a
discretised version of the inextensible WLC model. In this limit the
chain gets stiffer to stretching but keeps the same persistence
length. In this section we discuss how, in the case of $\nu
\rightarrow0$, the straightforward Euler method of integrating the
bead positions leads to an algorithm similar to the one developed by
Morse \cite{morse, montesi}.

For the limit $\nu \rightarrow 0$ at fixed $\kappa$ the time scaling
$\tau =\lambda t / \xi$ that we have been using so far, is not
useful. In this time scaling the coefficient of the stretching force
is set to unity. But now we employ a time scaling where $\lambda$ is
replaced by $\kappa$, such that the coefficient of the bending forces
equals unity.  The corresponding transformation, which we indicate
with a bar over the new variables, reads
\begin{equation} 
 \bar{\tau}=\kappa t /\xi = \nu \tau , \quad\quad \bar{T}^*= T^*/\nu, 
\quad \quad \bar{\bf g}_n = {\bf g}_n \nu,
 \label{y2}
\end{equation}
The reduced Hamiltonian then obtains the form
\begin{equation} 
  \bar{\cal H}^* = \left[ \frac{1}{2\nu}\sum^N_{n=1} 
    (|{\bf u}_n|-1)^2 - \sum^{N-1}_{n=1} {\bf u}_n \cdot {\bf
      u}_{n+1}\right].
\label{y3}
\end{equation} 
In terms of these new variables the dynamic equations change to
\begin{equation} \label{y4}
  \frac{d {\bf r}_n}{d \bar{\tau}} = - \frac{\partial \bar{\cal H}^*}{\partial {\bf r}_n} +\bar{\bf g}_n
\end{equation} 
with the correlation function
\begin{equation} \label{y5}
  \langle \bar{g}^\alpha_m (\bar{\tau})  \bar{g}^\beta_n (\bar{\tau}') \rangle = 2 \, \bar{T}^*
  \delta^{\alpha, \beta} \delta_{m,n} \delta(\bar{\tau}-\bar{\tau}').
\end{equation} 

In order to implement the limit $\nu \rightarrow 0$, we symbolically
write the equations of motion for the beads as
\begin{equation} 
  \frac{d {\bf r}_n}{d \bar{\tau}} = {\bf  f}_n= 
  - \frac{\partial \bar{\cal H}^*}{\partial {\bf r}_n} + \bar{\bf g}_n =
  -\frac{\partial \bar{\cal H}^*} {\partial {\bf u}_n}+
  \frac{\partial \bar{\cal H}^*} {\partial {\bf u}_{n+1} } + \bar{\bf g}_n.
  \label{y6}
\end{equation}
The differentiation in Eq. (\ref{y6}) is straightforward, with
exception of the first term in Eq. (\ref{y3}) which we write as
\begin{equation}
  \frac{\partial }{\partial{\bf u}_n}\frac{1}{2 \nu} 
  \left(\sum^N_{m=1} (|{\bf u}_m|-1)^2 \right)  \equiv  f^{\nu}_n \,{\bf u}_n.
 \label{y7}
\end{equation}  
The quantity $f^{\nu}_n$ may be considered as a ``tension'' that keeps
the length of the $n$-th bond close to unity. It obtains a finite
limit for $\nu \rightarrow 0$. Once we have an expression for the
$f^{\nu}_n$, the dynamical equations follow. We determine the value of
$f^{\nu}_n$ by the requirement that it keeps the evolution of the
chain configuration on the constrained subspace $u_n=1$; i.e.,
$f^{\nu}_n$ play the same role as the Lagrange multipliers that
preserve the contour length of the chain at all times in the
discretised version of the WLC \cite{morse}.

Since the $f^{\nu}_n$ are still undetermined we write the equations of
motion (\ref{y7}) as
\begin{equation} 
  \frac{d {\bf r}_n}{d \bar{\tau}} = -f^{\nu}_n \,{\bf u}_n + f^{\nu}_{n+1} {\bf u}_{n+1}  + {\bf f}^r_n,
  \label{y8}
\end{equation} 
where ${\bf f}^r_n$ contains the regular terms of the forces. Let us
then consider a finite increment $\Delta \tau$ in time. In this time
interval the bond vector ${\bf u}_n$ changes by the amount
\begin{equation} 
  \Delta {\bf u}_n = \left(f^{\nu}_{n-1} {\bf u}_{n-1} - 2 f^{\nu}_n \, {\bf u}_n + 
    f^{\nu}_{n+1}{\bf u}_{n+1} + {\bf f}^r_n -{\bf f}^r_{n-1} \right)  \Delta \tau,
  \label{y9}
\end{equation} 
leading to the tentative new value of the bond vector as
\begin{equation} 
  {\bf u}'_n = {\bf u}_n + \Delta {\bf u}_n.
  \label{y10}
\end{equation} 
Now requiring that $u'_n=u_n=1$ leads to the equations
\begin{equation} 
  2 {\bf u}_n \cdot \Delta {\bf u}_n +\Delta {\bf u}_n \cdot \Delta {\bf u}_n =0 \quad \quad
  {\rm or} \quad \quad ( {\bf u}_n + {\bf u}'_n )\cdot \Delta {\bf u}_n
  = 0.
\label{y11}
\end{equation} 
As we do not {\it a priori\/} know ${\bf u}'_n$, we first take as
zeroth order approximation ${\bf u}'_n={\bf u}_n$ and solve for
$\Delta {\bf u}_n$ in the second equation (\ref{y11}). Then we compute
the first approximation to ${\bf u}'_n$ with equation (\ref{y10}) and
iterate the cycle till it converges, which is typically reached in two
or three steps.

The structure of the second equation (\ref{y11}) is
\begin{equation} 
\begin{array}{rcl}
\left( \begin{array}{rcclcc}
-2 & b_1 & 0 & 0 & \cdots & 0\\*[2mm]
b_1 & -2 & b_2 & 0 & \cdots & 0 \\*[2mm]
0 & b_2 & -2 & b_3 & \cdots & 0 \\*[2mm]
\cdots &\cdots & \cdots &\cdots &\cdots & \cdots\\*[2mm]
0 & \cdots & 0 & b_{N-2} & -2 & b_{N-1} \\*[2mm]
0 & \cdots & 0 & 0 & b_{N-1} & -2 \\*[2mm]
\end{array}\right) 
\left( \begin{array}{c}
f^{\nu}_1 \\*[2mm] f^{\nu}_2 \\*[2mm] f^{\nu}_3\\*[2mm] \cdots \\*[2mm] f^{\nu}_{N-1} \\*[2mm] f^{\nu}_N \\*[2mm]
\end{array} \right) & = & \left( \begin{array}{c}
d_1 \\*[2mm] d_2 \\*[2mm] d_3\\*[2mm] \cdots \\*[2mm] d_{N-1} \\*[2mm] d_N \\*[2mm]
\end{array} \right),
\end{array}
\label{y12}
\end{equation}
with 
\begin{equation} 
  b_n = {\bf u}_n \cdot {\bf u}_{n+1} , \quad \quad \quad {\rm and} \quad \quad \quad
  d_n = ({\bf u}_n + {\bf u}'_n )\cdot ({\bf f}^r_{n-1}- {\bf f}^r_n).
\label{y13}
\end{equation} 
As the matrix in Eq. (\ref{y12}) is tridiagonal, the solution
$f^{\nu}_n$ is obtained by an ${\cal O} (N)$ operation. With the
converged $\Delta {\bf u}_n$ we update the bond vectors (which is
equivalent to updating the positions, since the centre-of-mass of the
chain is not affected by the motion).

This implementation of the inextensible WLC is an alternative for the
standard procedure of implementing the constraints \cite{morse} using
Lagrange multipliers that preserve the contour lengths of the chain at
all times. Not only that the parameters $f^{\nu}_n$ play the same role
as the Lagrange multipliers, but also the equations for the
$f^{\nu}_n$ are similar to the ones for the Lagrange parameters,
involving the same matrix as in Eq. (\ref{y12}). The difference is in
the right hand side of Eq. (\ref{y13}) and the definition of the
remaining forces ${\bf f}^r_n$ which involve in the standard procedure
additional metric pseudo-forces. Moreover, if we systematically
evaluate the forces at the midpoint
\begin{equation} 
  {\bf u}^m_n= ( {\bf u}_n + {\bf u}'_n )/| {\bf u}_n +
  {\bf u}'_n |,
  \label{y14}
\end{equation} 
then this scheme is symmetric in time between forward and backward
motion, which implies that detailed balance is obeyed to third order 
in the displacements.

\section{$\Delta\tau_{\text{max}}$ values for double-stranded
  DNA\label{sec6}}

\subsection{Time step for the inextensible WLC\label{inextens}}

For a fair comparison between the maximum allowable time-step
$\Delta\bar\tau_{\text{max}}$ between our model and the inextensible
WLC we have simulated our model at fixed values of $l_p/a$ (or
$\kappa$) for a series of decreasing values of the parameters $\nu$.
In order to stay close to dsDNA we have taken the value of $l_p/a$ of
dsDNA.  For the chain length $N=63$ beads we use the (default) Euler
scheme.

Recall from Sec. \ref{sec3a} that for $\nu=0.35$ nm the safe limit for
the Euler scheme for the bead position updates is $\Delta\tau=0.1$,
and the code becomes even unstable at $\Delta\tau\approx0.3$. Such
instabilities also occur for other values of $\nu$, and with
progressively smaller values of $\nu$ we found the stability limit to
behave as $\Delta \bar{\tau} \simeq 0.6 \nu$ --- the $\nu$ dependence
comes from the factor $1/ \nu$ in the harmonic confining potential
[the first term in the Hamiltonian (\ref{y3})]. So if we were to study
the maximum allowable time step from a series for decreasing $\nu$, we
would end up with time step zero for in the limit $\nu \rightarrow 0$.

The limit procedure as developed in Sec. \ref{limit} instead leads to
a finite allowable timestep, which thus is the largest that can be
used for small values of $\nu$ or in the inextensible limit. We find
that a time step $\Delta \bar{\tau} = 0.00005$ gets the equilibrium
average end-to-end distance and the average of the squared
displacement of the middle monomer within 5\% of the theoretically
calculated values. Translating this value to the scaling used by
Obermayer and Frey \cite{ober} (who implemented Morse's algorithm
\cite{morse,montesi}, and used the coefficient of the fluctuations of
the random forces to be equal to 2 as opposed to $2 \bar{T}^*$ as we
have used) we get a value $5\times10^{-6}$ for the time-step, which is
of the same order as used by them.

From this analysis one sees that using the forward Euler scheme,
modelling dsDNA as an (extensible) bead-spring model (that leads to
$\nu=0.35$) we get to $\Delta\tau_{\text{max}}=0.1$, which translates
to $\Delta\bar\tau_{\text{max}}\approx0.035$, while the same scheme
for inextensible WLC leads to
$\Delta\bar\tau_{\text{max}}=0.00005$. {\it I.e., just by allowing the
  chain to be naturally extensible into account we gain a factor $\sim
  10^3$ in the time-step.}

\subsection{Translating $\Delta\tau_{\text{max}}$ to real  times\label{sec7b}}

We now translate $\Delta\tau_{\text{max}}$ to real times using the
experimental parameters characteristic for dsDNA.

To this end we note that in Langevin dynamics there are no
hydrodynamic interactions among the beads, leading to the
centre-of-mass diffusion of a single chain $D=k_BT/(N\xi)$.  In
experiments hydrodynamic interactions are always present, however,
they only become important when the chain is long enough to exhibit
self-avoiding walk statistics. Thus, as long as the chains are
substantially smaller than the persistence length, they behave
essentially as straight rods, for which hydrodynamic interactions
among beads are not important. In other words, for chains
substantially smaller than the persistence length we can meaningfully
compare the centre-of-mass diffusion coefficient resulting from our
model and experiments. This comparison then yields us the
correspondence of $\Delta\tau_{\text{max}}$ to real times.

The diffusion coefficient of small dsDNA segments has been studied
using various techniques, such as capillary electrophoresis
\cite{stellwagena,stellwagenb}, dynamic light scattering \cite{dls},
NMR \cite{nmr}, and fluorescent recovery after photobleaching
\cite{frap}. Of these, the first three converge on the value $D\approx
1.07 \times10^8$ nm$^2$/s for a 20 bp dsDNA in water at room
temperature ($23^\circ$ C), while the last one reports $5.3\times10^7$
nm$^2$/s for dsDNA segment of length 21 bp. We decide to stick to the
values reported by the first three because of the consistency among
different experimental methods, and upon equating $D=k_BT/(N\xi)$ to
$\approx 1.07 \times10^8$ nm$^2/$s for $N=20$, with $k_BT=4.089$ pN nm
at $23^\circ$ C, we obtain
\begin{eqnarray} \label{xi}
\xi=1.91 \times10^{-12} \mbox{\,kg/s}.
\label{g1}
\end{eqnarray}
Further, writing the relation $t=\xi\tau/\lambda$ [see
Sec. \ref{sec2}, and the paragraph above Eq. (\ref{b3})], in terms of the 
parameters $T^*$ and $\nu$, the
conversion between the real time $t$ and the dimensionless time $\tau$
is obtained as
\begin{eqnarray}\label{time}
t=\frac{T^*}{b^2} \, \frac{a^2 \xi}{k_B T}  \, \tau =\nu\, \frac{a}{l_p} \, 
\frac{a^2 \xi}{k_B T}  \, \tau.
\label{g2}
\end{eqnarray}
The combination $a^2 \xi/(k_B T)$ is with $a=0.33$ nm and the value of
$\xi$ from Eq. (\ref{xi}) equal to
\begin{equation} \label{factor}
\frac{a^2 \xi}{k_B T} = 52.0 \mbox{\, ps}.
\end{equation}  
Inserting the dsDNA values $l_p/a=114$ and $\nu=0.35$ one gets 
\begin{equation} 
t \equiv c \tau, \quad \quad  {\rm with} \quad \quad c=0.16  \mbox{\,
  ps}.
\label{relat}
\end{equation} 
{i.e. \it one unit of dimensionless time 
  corresponds to $0.16$  ps of real time}. For the sake of
completeness, the corresponding conversion factor between $t$ and
$\bar\tau$ is given by
\begin{equation} 
t \equiv\bar c\bar\tau, \quad \quad  {\rm with} \quad \quad \bar c=0.45 \mbox{\,
  ps}.
\label{relata}
\end{equation} 
\begin{table*}
\begin{center}
\begin{tabular}{c||c||c||c}
  \hline\hline$\quad\mbox{chain length (bp)}\quad$&$\quad\Delta\tau_{\text{max}}\quad$
  &$\quad\Delta t_{\text{max}}$ (ps) $\quad$
  &$\quad t_{\text{max}}$ (ms) $\quad$ \tabularnewline\hline\hline
  64&10&1.59&0.23\tabularnewline\hline
  128&50&7.96&0.56\tabularnewline\hline
  256&100&15.9&0.56\tabularnewline\hline
  512&100&15.9&0.25\tabularnewline\hline\hline
\end{tabular}
\end{center}
\caption{Time forward integration steps $\Delta t_{\text{max}}$ in
  real times for various chain lengths. Also noted 
  in the last column the amount of real time $t_{\text{max}}$ our 
  model can simulate on a standard linux desktop computer in one hour.
  \label{table2}}
\end{table*}

With the above information we can now translate
$\Delta\tau_{\text{max}}$, the maximum time step, to real times
$\Delta t_{\text{max}}$ in Table \ref{table2}. Also noted in the last
column of Table \ref{table2} is the amount of real time
$t_{\text{max}}$ our model can simulate in an hour on a standard linux
desktop computer.

For completeness we mention the conversion of $\bar{\tau}$ to real times,
which is useful for small $\nu$ as occurring in the coarse-grained representation
of the polymer. This conversion reads in analogy with (\ref{g2})
\begin{eqnarray} 
  t=\frac{\bar{T}^*}{b^2}\, \frac{a^2 \xi}{k_B T} \,\bar{\tau} = \frac{a}{l_p} \, 
\frac{a^2 \xi}{k_B T} \,\bar{\tau} .
\label{g2a}
\end{eqnarray} 
With $\Delta \bar{\tau}_{\text{max}}\approx0.00005$ we find $\Delta t
_{\text{max}}\approx0.02$ fs, for dsDNA in the inextensible WLC limit,
which is about 5-6 orders of magnitude smaller than those listed in
the second column of Table \ref{table2}.
\begin{table}[h]
\begin{center}
\begin{tabular}{c||c}
\hline \hline$k$ &$c_k=t/\tau$ in ps\tabularnewline\hline\hline
$\quad1\quad$  & $\quad1.54\quad$\tabularnewline\hline
$\quad2\quad$  & $\quad13.09\quad$\tabularnewline\hline
$\quad5\quad$  & $\quad 119.434 \quad$\tabularnewline\hline
$\quad10\quad$  & $\quad 511.327 \quad$\tabularnewline\hline
$\quad12\quad$  & $\quad 741.622 \quad$\tabularnewline\hline
$\quad15\quad$  & $\quad 1165.66 \quad$\tabularnewline\hline
$\quad20\quad$  &$\quad 2081.9  \quad$\tabularnewline\hline
\end{tabular}
\end{center}
\caption{The ratio between the real time $t$ and the scaled time 
  $\tau$ under coarse-graining, with coarse-graining parameter
  $k$. The corresponding values of $T^*_k$ and $\nu_k$ can be found in
  Table \ref{table3}.\label{table5}}
\end{table}

\begin{figure}[h]
\begin{center}
\includegraphics[width=0.6\linewidth]{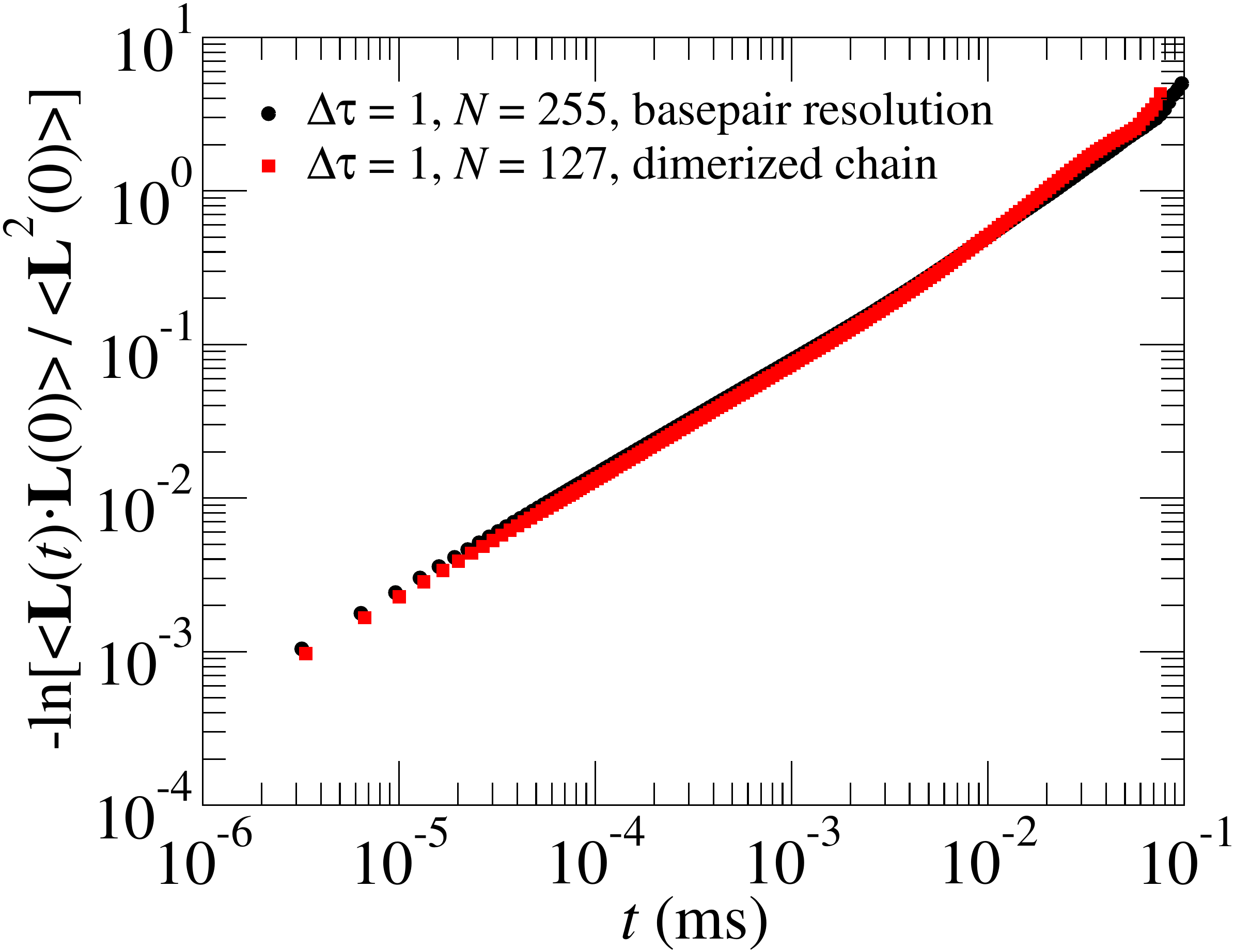}
\end{center}
\caption{(Color online) The end-to-end vector data for a chain of 256 
  base pairs and that of the corresponding chain of 128 dimerized 
  basepairs. \label{compar}}
\end{figure}

We also note that coarse-graining increases the factor between real
and scaled time substantially, as can be seen from Table \ref{table5}.
For the dimerized dsDNA chain we show in Fig. \ref{compar} the time
evolution of the MSD of the end-to-end vector for a chain of 256 base
pairs and that of 128 dimerized base pairs. Both have the same
force-extension curves and the latter is simulated with the parameters
$T^*_2=0.008$ and $\nu_2=0.1875$.  The figure shows that the
correspondence between the two is excellent. In general, the net gain
in real time upon coarse-graining remains modest: although the value
of $c_k$ increases with $k$ as listed in Table \ref{table5}, it also
leads to smaller $\nu_k$, further leading to smaller allowable time
steps, as discussed in Sec. \ref{inextens}. The eventual largest
time-step under coarse-graining for dsDNA, in real time, is still
larger than the allowable time step for the $\nu=0$ case for
inextensible WLC; i.e., even with a small $\nu_k$ it remains efficient
to use the model rather than the limit $\nu=0$ procedure.  A further
advantage of coarse-graining is that longer polymers can be simulated
in the time associated with the smaller number of beads, but that is
true for any model.
 
\section{Conclusion\label{sec7}}

Using a recently developed bead-spring model, in the absence of
hydrodynamic interactions among the beads, in this paper we have
developed an efficient algorithm to simulate the dynamics of dsDNA as
a semiflexible polymer. Polymer dynamics in the model is described by
the Langevin equation. We consider dsDNA at persistence length
$l_p\approx37.7$ nm that corresponds to $114$ beads in the model. The
model can be mapped one-on-one to the extensible WLC. We show that,
within an accuracy tolerance level of $5\%$ of several key
observables, the model allows for large single Langevin time steps; as
summarised in Table \ref{table2}.

The key to such large time steps is to use the polymer's fluctuation
modes as opposed to the individual beads for simulating the
dynamics. The conventional simulation approach would be to integrate
the corresponding Langevin equations of motion in time, with a simple
integration scheme such as the Euler method. This is however, not an
efficient method, as one can get to $\Delta t_{\text{max}}$ to
$\approx0.16$ ps; at $\Delta t_{\text{max}}\approx 0.48$ ps the
integration algorithm even becomes unstable.  Instead, we use the
polymer's fluctuation modes to integrate the dynamical equations
forward in time. Although any choice of orthogonal basis functions can
be used to describe the polymer's fluctuation modes, we found that the
choice of the Rouse modes provides the most stable and robust
results. Use of the Rouse modes allows us to take 2 to 3 orders of
magnitude larger time steps for integrating the Langevin equations
forward in time, as evidenced in Table \ref{table2}.

We remark that the numbers in the table are only indicative for the
order of magnitude of the allowable time step for various
reasons. First of all, the choice 5\% for the equilibrium quantities
(consistently, 10\% for the dynamical quantities) is arbitrary.
Secondly, the percentage error in the physical observables depends on
the quantity chosen. We find that in most cases the msd of the middle
monomer decides the size of $\Delta t_{\text{max}}$. The end-to-end
vector also plays that role in a few cases, while the middle bond is
not at all critical. For the accuracy percentage (of course) it also
matters whether one takes e.g. the square of the end-to-end vector (as
we did) or the vector itself (the latter choice halves the
error). Finally, the maximum allowable time step depends also on the
parameters $T^*$ and $\nu$. We found that the dependence on $T^*$ is
rather weak but the sensitivity to the value of $\nu$ is much
stronger. Nevertheless, despite these reservations, the gain of 2 to 3
orders of magnitude by changing the integration variables from bead
positions to Rouse modes stands firm. Importantly, a similar speed-up
can also be achieved for the extensible WLC, since our model can be
mapped one-on-one to it.

Like almost any other model, ours allows for coarse-graining. As the
model is restricted to conform to the force-extension curve, the two
parameters of the model $T^*$ and $\nu$ travels through a sequence of
points, all leading to the same force-extension curve. We have shown
that $\nu$ becomes progressively smaller under increasing degree of
coarse-graining. Although not really applicable to the case for dsDNA
(but certainly for f-actin), $\nu$ can be made to become so small that
physically our model approaches the limit of inextensible WLC. We have
shown that in the limit of $\nu\rightarrow0$ the dynamical equations
of the beads approach a form similar to the ones developed by Morse
\cite{morse,montesi} for simulating the inextensible WLC. Using these
equations we have simulated a dsDNA chain of length $N=63$ in the
inextensible limit using the bead positions as dynamical variables
(with the default Euler updating scheme). By means of doing so,we have
demonstrated that just by changing the model from inextensible to
extensible bead-spring we gain $\approx 3$ orders of magnitude in the
size of the time-step. Combining this speed-up with the ones achieved
by using the Rouse modes as integration variables as opposed to the
bead positions, we have achieved 5-6 orders of magnitude speed-up in
the size of the time-step in comparison to the inextensible WLC.

Further, we note that Langevin dynamics simulations are widely used
for the simulation of biopolymers, but most publications do not
present a clear translation of the simulation time in real time
(picoseconds or nanoseconds) and the experimental observation used to
make this translation; and certainly do not explore the maximal time
step which does not cause significant systematic errors. We found
reports of time steps of 12.9 ps \cite{timestep1}, 5 ps
\cite{timestep2} and 3.8 ps \cite{timestep3}, in simulations in which
a bead represents 4, 9 and 37 basepairs respectively; A
back-of-the-envelop estimate then yields that by-and-large, these time
steps are below the maximal time step estimated by us for an
integration scheme in real-space coordinates.

Finally, while our model is not well-suited for hard-core interactions,
it does allow for adding other forces to the monomers. In
order to showcase this, we have simulated dsDNA segments in a shear
flow, where the viscous drag force due to the shear flow makes the
chain tumble in space. The equations of motion Eq.~(\ref{b1}) then become
\begin{eqnarray}
\frac{d {\bf r}_n (t)}{d t } = -\frac1\xi\frac{\partial {\cal H}}{\partial {\bf r}_n}
  + \dot{\gamma}\, y_n \hat{\bf x} + {\bf g}_n (t),
\label{btum}
\end{eqnarray}
where $\dot{\gamma}$ is the shear rate.
These equations are easily transformed into mode equations.
In the Supplementary Information we present a
movie of a tumbling dsDNA chain of 255 base pairs; the chain tumbles
in water with a velocity field ${\bf v}({\bf r})=\dot\gamma y\hat x$,
with shear rate $\dot\gamma\approx1.54\times10^8$ s$^{-1}$ (which
corresponds to Weissenberg number $\mbox{Wi}\approx5.55\times10^3$,
calculated from the moment of inertia of a straight rod of the same
length as the dsDNA segment \cite{leeuwen,leeuwen1}). In the movie the
centre-of-mass of the chain always remains at the origin of the
co-ordinate system. The data for the movie are generated with $\Delta
t=\Delta t_{\text{max}}=16$ ps, and took about three minutes to
generate on a linux desktop; it contains 3,000 snapshots, with
consecutive snapshots being 8 ns apart. A detailed study of the
tumbling motion of the dsDNA in a shear flow is however not the focus
of this paper; it will be taken up in an upcoming one.

\section*{Acknowledgements}

Ample computer time from the Dutch national cluster SARA is gratefully
acknowledged. 
\appendix

\section{Coarse-graining f-actin}

The persistence length of f-actin is orders of magnitude longer than
that of dsDNA. Liu and Pollack \cite{liu} report values
$l_p=8.75\,\mu$m for f-actin, while the length of actin monomer is 5.5
nm \cite{stef}. Also the force constant $K_0$ is orders larger than
the dsDNA value. Liu and Pollack find for $K_0= 35.5$ nN.  This leads
to the dimensionless parameters
\begin{equation} 
  r = l_p/a=1591 \quad \mbox{and} \quad y = 7.65\times10^7
\label{A1}
\end{equation} 
for f-actin. In order to determine $\xi$, we use the information that
the time taken for an actin filament with a contour length $L=10$
microns (corresponding to $N=L/a=1818$ beads) and a diameter of 5 nm
in a solution with a viscosity of 0.1 Pa s (water) at a temperature of
$20^\circ$C has to diffuse its own length is $t^*=1.5\times10^4$
s. This leads us to the equation
\begin{equation} 
  t^*=L^2/(6D)=\xi L^3/(6a
  k_BT);\quad\mbox{i.e.,}\quad\xi=5.5\times10^{-11}\,\,\mbox{kg/s},
\label{A1a}
\end{equation} 
where $D$ denotes the diffusion coefficient of the f-actin filament,
further yielding 
\begin{equation} 
  \frac{a^2\xi}{k_BT}\approx0.4\,\,\mu\mbox{s}.
\label{A1b}
\end{equation} 

With these values we can now calculate with the expressions (\ref{x3})
the values of the effective $\bar T^*_k$ and $\nu_k$ for f-actin,
which are shown in Table \ref{table4}.
\begin{table}[h]
\begin{center}
\begin{tabular}{c||c||c||c} 
\hline \hline 
\quad$k$\quad  & $\bar T^*_k$ & $\nu_k$&$\bar c_k$ (in $\mu$s)\tabularnewline \hline\hline
\quad1\quad   & \quad 0.000713 \quad  & \quad   0.030603\quad &\quad 0.00025\quad  \tabularnewline \hline
\quad2\quad   & \quad 0.001298 \quad & \quad   0.008019\quad &\quad 0.00402\quad    \tabularnewline \hline
\quad5\quad   & \quad 0.003159 \quad & \quad   0.001301 \quad &\quad 0.15713\quad   \tabularnewline \hline
\quad10\quad   &  \quad 0.006294 \quad &\quad 0.000326 \quad &\quad 2.51414\quad  \tabularnewline \hline
\quad20\quad   & \quad  0.012575 \quad & \quad 8.14832 $\times10^{-5}$ \quad &\quad 40.2263 \quad   \tabularnewline \hline
\quad50 \quad  & \quad  0.031428  \quad&  \quad 1.30391 $\times10^{-5}$\quad &\quad 1571.34 \quad   \tabularnewline \hline
\hline 
\end{tabular}
\end{center}
 \caption{Coarse-grained parameters for f-actin. \label{table4}}
\end{table}

The variation in $\bar T^*_k$ with $k$ does not have significant
consequences since $\bar T^*_k$ enters the dynamical equations only in
the form of $\sqrt{\bar T^*_k}$. The decrease of $\nu_k$ with
increasing $k$, on the other hand, has much more severe consequences
on the dynamics, as it makes the chain effectively more
inextensible. In contrast to dsDNA one already runs into --- even for
fairly mild coarse-graining --- quite small values of $\nu_k$. E.g.,
for $k=50$, which would mean about 75 beads per persistence length,
the value of $\nu_k$ is so small that for all practical purposes our
model behaves (except of course the force-extension curve) as the
inextensible WLC, and our maximum time-step then would then be
$\Delta\bar\tau_{\text{max}}\approx0.00005$.

Even more interesting is the ratio between the real time and the
scaled time involving the $k$-dependent parameters
[c.f. Eq. (\ref{g2a})]:
\begin{equation} 
  t =\bar c_k \bar{\tau} = k^4 \frac{a}{l_p} \frac{a^2 \xi}{k_B T}
  \bar{\tau}.
\label{A2}
\end{equation} 
E.g., even with a small maximal allowable timestep $\Delta
\bar{\tau}_{\text{max}} = 0.00005$ the large value of $\bar c_k$ for
$k=50$ leads to
\begin{equation} 
\Delta t_{\text{max}} = 0.00005\times1571.34\times0.4 \,\mu{\rm s} = 0.03 \,\mu{\rm
  s}.
\label{A3}
\end{equation} 
This is orders of magnitude larger than the order of picosecond
estimates for $\Delta t_{\text{max}} $ for dsDNA.

\end{document}